\documentclass[useAMS,usegraphicx,usenatbib]{mn2e}
\usepackage{multirow}
\usepackage{amssymb}

\newcommand{\kms}{\mbox{$\>{\rm km\, s^{-1}}$}}

\newcommand{\Mpc}{\>{\rm Mpc}}
\newcommand{\kpc}{\mbox{$\>{\rm kpc}$}} 
\newcommand{\Gyr}{\mbox{$\>{\rm Gyr}$}}
\newcommand{\msun}{\>{\rm M_{\odot}}}

\newcommand\degrees{^\circ}
\newcommand{\Mdisc}{\mbox{$M_{\rm d}$}}
\newcommand{\Rd}{\mbox{$R_{\rm d}$}}
\newcommand{\zd}{\mbox{$z_{\rm d}$}}
 
\def\etal{{et al.}}

\def\ie{{\it i.e.}}

\title[Red versus blue discs in haloes]{Internal alignments of red
  versus blue discs in dark matter haloes}
   
\author[Debattista \etal]{Victor P. Debattista$^1$\thanks{E-mail:
    vpdebattista@gmail.com}, Frank C. van den Bosch$^2$, Rok
  Ro\v{s}kar$^3$,
  Thomas Quinn$^4$, \newauthor  Ben Moore$^3$, David R. Cole$^1$ \\
  $^1$ Jeremiah Horrocks Institute, University of Central Lancashire,
  Preston, PR1 2HE, UK \\
  $^2$ Astronomy Department, Yale University, PO Box 208101,
  New Haven, CT 06520-8101, USA \\
  $^3$ Research Informatics, Scientific IT Services, ETH Z\"urich,
  Weinbergstrasse 11, CH-8092, Z\"urich, Switzerland \\
  $^4$ Astronomy Department, University of Washington, Box 351580,
  Seattle, WA 98195, USA \\
}

\begin{document}   

\date{{\it Draft version on \today}}
\pagerange{\pageref{firstpage}--\pageref{lastpage}} \pubyear{----}
\maketitle

\label{firstpage}

\begin{abstract} 
  Large surveys have shown that red galaxies are preferentially
  aligned with their haloes, while blue galaxies have a more isotropic
  distribution.  Since haloes generally align with their filaments, this
  introduces a bias in the measurement of the cosmic shear from weak
  lensing.  It is therefore vitally important to understand why this
  difference arises.  We explore the stability of different disc
  orientations within triaxial haloes.  We show that, in the absence of
  gas, the disc orientation is most stable when its spin is along the
  minor axis of the halo.
  Instead when gas cools on to a disc it is able to form in almost
  arbitrary orientation, including off the main planes of the halo
  (but avoiding an orientation perpendicular to the halo's
  intermediate axis).
  Substructure helps gasless galaxies reach alignment with the halo
  faster, but has less effect on galaxies when gas is cooling on to
  the disc.
  Our results provide a novel and natural interpretation for why red,
  gas poor galaxies are preferentially aligned with their halo, while
  blue, star-forming, galaxies have nearly random orientations, without
  requiring a connection between galaxies' current star formation rate
  and their merger history.
\end{abstract}


%
%

\section{Introduction}
\label{sec:intro}

Intrinsic alignments of galaxies, i.e., correlations of galaxy shapes
with each other or with the large scale density field, are one of the
most important astrophysical contaminants for weak lensing
measurements \citep[see][for a comprehensive review]{troxel_ishak15}.
In particular, coherent alignments of the shapes of physically nearby
galaxies can strongly bias the cosmic shear measurements that lie at
the heart of upcoming experiments such as the Large Synoptic Survey
Telescope \citep{lsst}, Euclid \citep{euclid}, the Dark Energy Survey
\citep{des} and the {\it Wide-Field Infrared Survey Telescope}
\citep{wfirst}. If ignored, this intrinsic alignment bias can cause
large systematic errors in, for example, the inferred dark energy
equation of state \citep[e.g.][]{bridle_king07, kirk+12}. It is
prudent, therefore, that we develop a solid understanding of galaxy
alignments.

Observationally, various forms of alignment have been detected over a
wide range of scales.  On scales ranging from $\sim 1$ to $100 h^{-1}
\Mpc$, numerous studies have detected alignments between individual
clusters \citep[e.g.][]{binggeli82, plionis94, smargon+12}, between
galaxies \citep[e.g.][]{pen+00, brown+02, heymans+04, lee_pen07,
  okumura+09, li+13}, and between galaxies and their surrounding
large-scale density (or tidal) field \citep[e.g.][]{mandelbaum+06,
  hirata+07, paz+08, faltenbacher+09, tempel_libeskind13, tempel+13,
  zhangy+13, zhangy+15}. Alignments have also been detected on smaller
scales, within individual host haloes. These include, among others,
the alignment between brightest cluster galaxies and their parent
cluster \citep[e.g.][]{carter_metcalfe80, binggeli82, struble90},
between the orientation of central galaxies and the spatial
distribution of their satellite galaxies
\citep[e.g.][]{sales_lambas04, brainerd05, yang+06,
  agustsson_brainerd06, azzaro+07, faltenbacher+07, wangy+08,
  wangy+08b, wangy+10, nierenberg+11, li+13}, as well as between the
orientation of satellite galaxies and the central-satellite position
vector \citep[e.g.][]{pereira_kuhn05, agustsson_brainerd06,
  faltenbacher+07, hao+11, schneider+13}.  An important outcome from
all these studies is that the strength of these various alignments
depends on a variety of galaxy properties. Of particular relevance to
this paper is the fact that alignment effects are typically much
stronger for red (early-type) galaxies than for blue (spiral)
galaxies.

From the theoretical side, tidal torque theory predicts large-scale
correlations between the angular momenta and shapes of dark matter
haloes \citep[e.g.][]{catelan_porciani01, catelan+01, crittenden+01,
  porciani+02}, which has been tested and confirmed with numerical
simulations \citep[e.g.][]{croft_metzler00, heavens+00, jing02,
  faltenbacher+08}. The tidal field is however too weak to directly
affect the orientation of galaxies and must do so either through
formation or accretion processes \citep{camelio_lombardi15}.
Simulations have also shown that the shapes and/or spins of dark
matter haloes are aligned with their large-scale density distribution
\citep[e.g.][]{aragon-calvo+07, hahn+07a, hahn+07b, cuesta+08, paz+08,
  zhangy+09, wang+11, codis+12, libeskind+12, trowland+13}. Such
alignments originate from both tidal torques, and from the preferred
accretion of new material along the directions of sheets and filaments
that delineate the cosmic web \citep[e.g.][]{jing02,
  bailin_steinmetz05, faltenbacher+05}.  On small scales, inside the
virialized regions of dark matter haloes, non-linear effects such as
violent relaxation and impulsive encounters are believed to weaken any
pre-infall alignments \citep[e.g.][]{porciani+02}.  However, at the
same time tidal forces due to the host halo may introduce new
alignments, similar to the tidal-locking mechanism that affects the
Earth-Moon system \citep[e.g.][]{ciotti_dutta94, usami_fujimoto97,
  fleck_kuhn03}.  Indeed, numerical simulations have shown that dark
matter subhaloes are preferentially radially aligned, with their major
axes pointing towards the centre of their host halo
\citep[e.g.][]{kuhlen+07, faltenbacher+08, pereira+08}.

Hence, if {\it galaxy} shapes and/or orientations are determined
either by the angular momentum or by the shape/orientation of the {\it
  halo} in which they form, then intrinsic galaxy alignments will
naturally emerge.  The vast majority of galaxies, including blue,
spiral galaxies but also the majority of red, early-type systems, are
supported by rotation \citep[see e.g.,][]{emsellem+11}, and their
orientation is therefore governed by angular momentum.\footnote{Only
  the most massive ellipticals are pressure supported, and their
  orientations most likely reflect the directionality of their last
  major merger} Thus, understanding intrinsic alignments requires
understanding how the angular momentum vectors of discs align with
their host haloes and their large-scale tidal field. Collisionless
simulations find angular momenta of haloes aligned with their minor
axes, with only small misalignments \citep{dubinski92, war_etal_92,
  porciani+02, bailin_steinmetz05, faltenbacher+05}.  However,
hydrodynamical simulations reveal a more complicated picture. First of
all, even in the absence of cooling, the halo and gas angular momenta
misalign on average by $\sim 30\degrees$ \citep{vdb_etal_02,
  che_etal_03, sha_ste_05}. Furthermore, when cooling is included, the
formation of the disc modifies the shape and orientation of the inner
halo, but leaves the outer halo largely intact \citep[][hereafter
  D08]{kkzanm04, bailin+05, debattista+08}.  Consequently, the disc
angular momentum is well aligned with the inner halo ($r \la 0.1
r_{\rm vir}$), but is only poorly aligned with the minor axis at
larger halocentric radii; overall hydrodynamical simulations find that
discs and haloes are misaligned by $30\degrees$-$40\degrees$ on
average \citep{croft+09, bett+10, hahn+10}. This is of the same order
as what is required to reconcile the relatively weak alignment
strength observed, with the relatively strong alignments predicted
from pure dark matter simulations \citep[e.g.][]{kang+07, wangy+08,
  joachimi+13b, schneider+13, zhangy+15}.

However, what is the cause of the colour-dependence of intrinsic
alignment strength? It has been suggested \citep[e.g.][]{joachimi+13a,
  joachimi+13b} that blue disc galaxies are aligned with the angular
momentum vector of their host halo, while red, early-types have
moments of inertia that are aligned with those of their host haloes.
The latter is motivated by the belief that early-type galaxies are the
outcome of major mergers, combined with the idea that merger remnants
are oriented along the direction of the last major merger
\citep[e.g.][]{vanhaarlem_vandeweygaert93}.  However, as already
alluded to above, the majority of red, early-type galaxies are discs:
the ATLAS$^{\mathrm{3D}}$ project finds that $\sim 86\%$ of early-type
galaxies (defined as galaxies that lack spiral arms) are rapidly
rotating \citep{emsellem+11}. Hence, we consider it unlikely that the
origin of the different orientations of red versus blue galaxies is
predominantly accounted for by mergers.

In this paper, we examine an alternative explanation that accounts for
the fact that (i) dark matter haloes are triaxial \citep{bbks_86,
  bar_efs_87, frenk_etal_88, dub_car_91, jin_sut_02,
  bailin_steinmetz05, all_etal_06}, giving rise to torques between the
halo and the disc, (ii) dark matter haloes have substructure, which
can tilt discs \citep{huang_carlberg97} and cause gravitational
perturbations, and (iii) what distinguishes red from blue discs is the
lack of ongoing gas accretion.  In \citet[][hereafter
  D13]{debattista+13b}, we explored the stability of disc galaxies
with their spins along the intermediate axis of the halo, an
orientation that has been proposed for the Milky Way \citep{law+09,
  law_majewski10, deg_widrow13}.  D13 showed that such an orientation
is never stable, and discs can never form in this orientation even
when gas angular momentum is along the intermediate axis.  D13 also
showed that a disc can survive for a long time off one of the symmetry
planes of the halo when gas is present, and proposed this as a natural
interpretation for the Milky Way.  In this paper, we extend this
earlier work to address the stability of disc orientations in general,
and the role of gas in supporting different orientations.  We show
that gas cooling provides a dominant mechanism for maintaining discs
off the symmetry planes of the halo, and that this offers a natural
explanation for the different orientations of red and blue galaxies.
In particular, we suggest that red discs lack significant gas
accretion, and have their orientation governed by halo torques, while
blue discs have orientations that are set by the balance between halo
torques and angular momentum of ongoing gas accretion.


\section{Collisionless simulations}
\label{sec:collisionless}

We start by considering the evolution of disc galaxy orientation in
the absence of complications introduced by gas.

\subsection{Constructing collisionless initial conditions}

As in D08, we form triaxial haloes via the head-on merger of two
prolate haloes, themselves the product of head-on mergers of spherical
haloes \citep{moo_etal_04}.  The mergers, and all subsequent
collisionless simulations, are evolved with {\sc pkdgrav}
\citep{stadel_phd}, an efficient, multi-stepping, parallel tree code.

In the collisionless simulations, we use halo A of D08.  We have
verified that other halo models give similar results.  Halo A is
constructed from two consecutive mergers.  The first head-on merger
places two concentration ${\cal C}=10$ spherical haloes of mass $2.3
\times 10^{12}\msun$ 800 kpc apart approaching each other at 50~\kms.
The spherical haloes are generated from a distribution function using
the method of \citet{kazantzidis+04} with each halo composed of two
mass species arranged on shells.  The outer shell has more massive
particles than the inner one, increasing the effective resolution in
the centre.  We use a softening parameter $\epsilon = 0.1\kpc$
($\epsilon = 0.5\kpc$) for low (high) mass particles.  As shown in
D08, a large part of the particle mass segregation persists after the
mergers and the inner region remains dominated by low-mass particles.
This merger produces a prolate halo.  Halo A results from the head-on
merger of two copies of this prolate halo starting 400 kpc apart at
rest.  The progenitor spherical haloes each have 1 million particles,
equally divided between their two mass species.  The outer halo
particles are $\sim 19\times$ more massive than the inner ones.  Thus,
in total halo A consists of 4 million particles.  The shape of halo A
was presented in D13; Table \ref{tab:haloes} lists its
properties\footnote{We use a different convention for $r_{200}$ from
  D08, \citet{valluri+10} and \citet{valluri+12}, who used the radius
  {\it at} which $\rho = 200 \rho_{crit}$.  Here $r_{\rm 200}$ is the
  radius {\it within} which the enclosed mass has average density $200
  \rho_{crit}$.}.  It has $c/a \sim 0.35$, but only a mild triaxiality
\citep{fra_etal_91} $T = (a^2 - b^2)/(a^2 - c^2) \simeq 0.9$ out to
100 kpc.

\begin{centering}
\begin{table}
\vbox{\hfil
\begin{tabular}{cccccccc}\hline 
\multicolumn{1}{c}{Halo} &
\multicolumn{1}{c}{$N_p$} &
\multicolumn{1}{c}{$N_g$} &
\multicolumn{1}{c}{$M_{\rm 200}$} &
\multicolumn{1}{c}{$r_{\rm 200}$} &
\multicolumn{1}{c}{$b/a$} &
\multicolumn{1}{c}{$c/a$} \\

 & $(10^6)$ & $(10^6)$ & ($10^{12} \msun$) & (kpc) & & \\ \hline

A   & 3.3 & - & 6.3 & 379 & 0.45 & 0.35 \\ 
GP  & 1.4 & 1.3 & 1.6 & 239 & 0.55  & 0.55 \\ 
GT & 2.8 & 2.7 & 3.2 & 304 & 0.4   & 0.32 \\ \hline 
\end{tabular}
\hfil}
\caption{
The haloes used in the simulations of this paper.  The
properties listed are for the halo after the last merger and before
the discs have been grown.  $N_p$ and $N_g$ are the number of dark
matter and gas particles within $r_{\rm 200}$, and $M_{\rm 200}$ is
the halo mass within the virial radius, $r_{\rm 200}$.
Density axes-ratios $b/a$ and $c/a$ are by-eye averaged over the inner 
20 kpc.}
\label{tab:haloes}
\end{table}
\end{centering}

Once we produce the triaxial halo, we insert a disc of particles.  The
disc distribution is exponential with scalelength $\Rd = 3 \kpc$ in
all cases except model LA2d, which had $\Rd = 6 \kpc$, and Gaussian
scaleheight $z_{\rm d} = 0.05 R_{\rm d}$.  The discs are placed at
various orientations within the halo.  We refer to these experiments
by the halo axis along which the disc's spin is aligned: in
'short-axis' (S) experiments, the disc spin is along the short axis of
the halo, while in 'long-axis' (L) experiments, the disc spin is along
the halo's long axis.  An 'intermediate-axis' (I) experiment has the
disc spin along the halo's intermediate axis.  Most of the
intermediate-axis simulations were presented already in D13; here, we
include a new model, IA3, with a significantly more massive disc.  The
discs are comprised of $3\times 10^5$ equal-mass particles.  Initially
the disc has negligible mass, but we increase this adiabatically
linearly over time to a mass \Mdisc.  During this time, the halo
particles are free to move and remain in equilibrium while the disc
particles are held fixed in place.

\begin{table}
\begin{centering}
\begin{tabular}{cccc}\hline 
\multicolumn{1}{c}{Model} &
\multicolumn{1}{c}{\Mdisc} &
\multicolumn{1}{c}{\Rd} &
\multicolumn{1}{c}{\zd} \\

 & ($10^{10} \msun$) & (kpc) & (pc) \\ \hline

 SA1 &  5.2 & 3 & 150 \\ 
 IA1 &  17.5 & 3 & 150 \\ 
 IA2 &   7.0 & 3 & 150 \\ 
 IA3 &  70.0 & 3 & 150 \\ 
 LA1 &  3.5 & 3 & 150 \\ 
 LA2 &  10.5 & 3 & 150 \\ 
 LA2d &  10.5 & 6 & 300 \\ 
\hline
\end{tabular}
\caption{The collisionless simulations of this paper.  \Mdisc\ is the
  mass of the disc, \Rd\ is the disc exponential scalelength and \zd\
  is the Gaussian scaleheight of the disc.  All models are grown
  inside halo A.}
\label{tab:simulations}
\end{centering}
\end{table}

Once the disc reaches the target mass, we set its particle kinematics
appropriate for the constant disc-height $z_d$ and Toomre-$Q = 1.5$,
as described in \citet{debattista_sellwood00}.  For this, we calculate
the potential using a hybrid polar-grid code with the disc on a
cylindrical grid and the halo on a spherical grid \citep{sellwood03}.
In setting up the disc kinematics, we azimuthally average radial and
vertical forces.  From these initial conditions, the models are then
evolved with {\sc pkdgrav}, with timesteps refined such that $\delta t
= \Delta t/2^n < \eta (\epsilon/a_g)^{1/2}$, where $\epsilon$ is the
softening and $a_g$ is the acceleration at a particle's current
position.  We use base timestep $\Delta t = 5$ Myr in all cases except
model SA1 (which uses $\Delta t = 50$ Myr).  For all simulations, we
use $\eta = 0.2$ and an opening angle of the tree code $\theta = 0.7$.
Table \ref{tab:simulations} lists the collisionless simulations
discussed in this paper.\footnote{The names of models in this paper are
  different from those in D08.}

\subsection{Briggs figures}
\label{ssec:briggsfigs}

The orientation of a disc inside a triaxial halo is necessarily a 3D
property of a galaxy.  Consequently visualizing this in a single 2D
figure requires a projection and some loss of information.  Because we
are only interested in the orientation of the disc, which we quantify
via the orientation of the disc angular momentum, it is sufficient to
indicate the two orientation angles of the disc angular momentum.  The
lost information is the amplitude of the angular momentum vector, but
this quantity is of no interest here.  Briggs figures
\citep{briggs_90}, which are 2D polar coordinate representations of
the direction of vectors, are ideal for this purpose.  A Briggs figure
plots the standard two spherical angular coordinates $\theta$ and
$\phi$ as the radial and angle coordinates in 2D polar coordinates.

We use Briggs figures to present the evolution of the discs in our
simulations relative to the dark matter halo.  We hold the coordinate
frame with respect to which the angles $\theta$ and $\phi$ are defined
fixed, so changes in $\theta$ and $\phi$ are a result of disc tilting;
because our haloes do not tumble (by construction) the directions of
the principal axes of the halo, which we also indicate in the Briggs
figures, do not change.  Our long, short and intermediate axes are
always defined at the virial radius, not at small radii where the halo
shape is changed by the baryons.  As our stellar discs are not
strongly warped, we indicate only the orientation of the net angular
momentum of the stars out to a radius of 15~\kpc.

A note about how we indicate the principal axes in Briggs figures.  We
have chosen symbols in a simple visual mnemonic to indicate the short,
intermediate and long axes: throughout we use a triangle, square and
star, respectively to represent them.  Thus, the ordering of the axes
is the same as the ordering of the number of sides in their
corresponding symbols.

\begin{figure*}
\centerline{
\includegraphics[angle=0.,width=0.5\hsize]{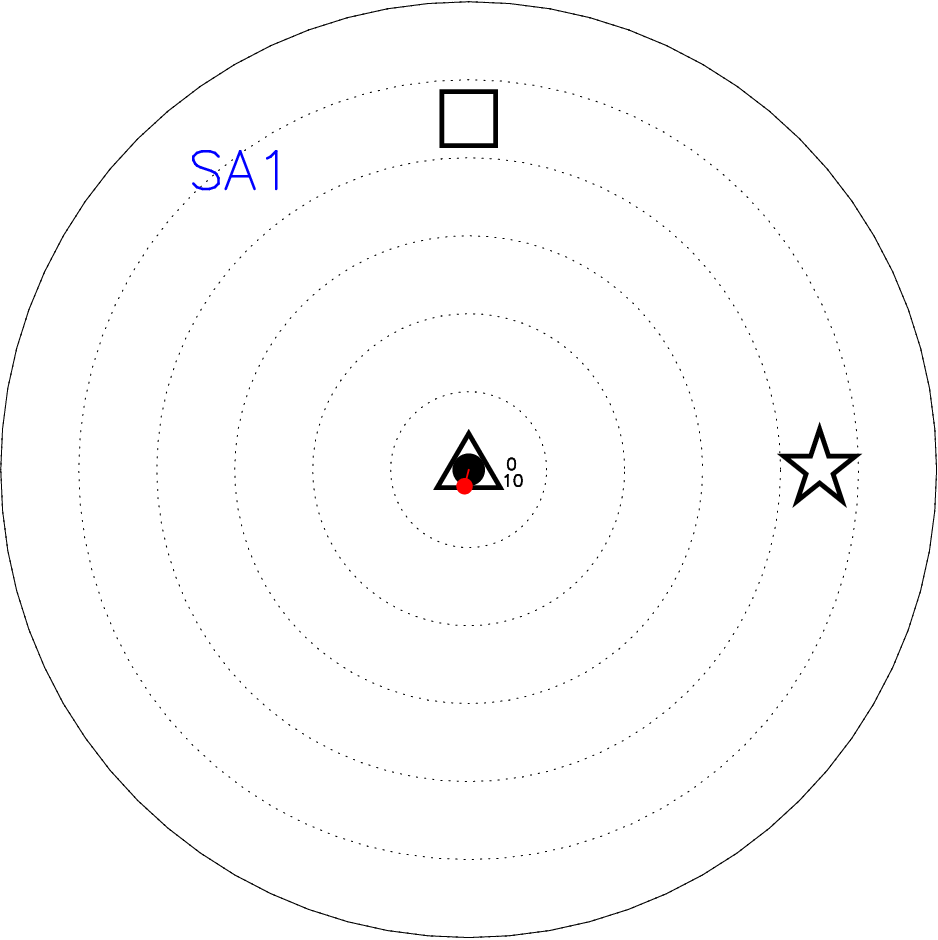}
\includegraphics[angle=0.,width=0.5\hsize]{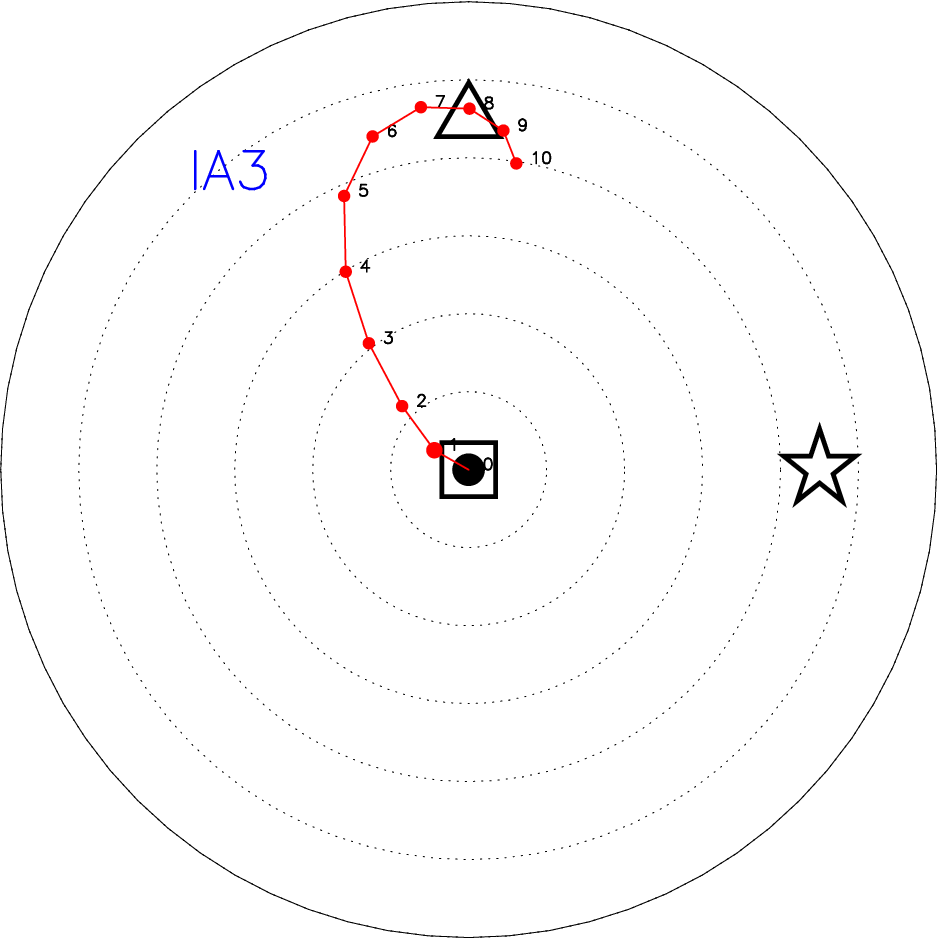}
}
\centerline{
\includegraphics[angle=0.,width=0.5\hsize]{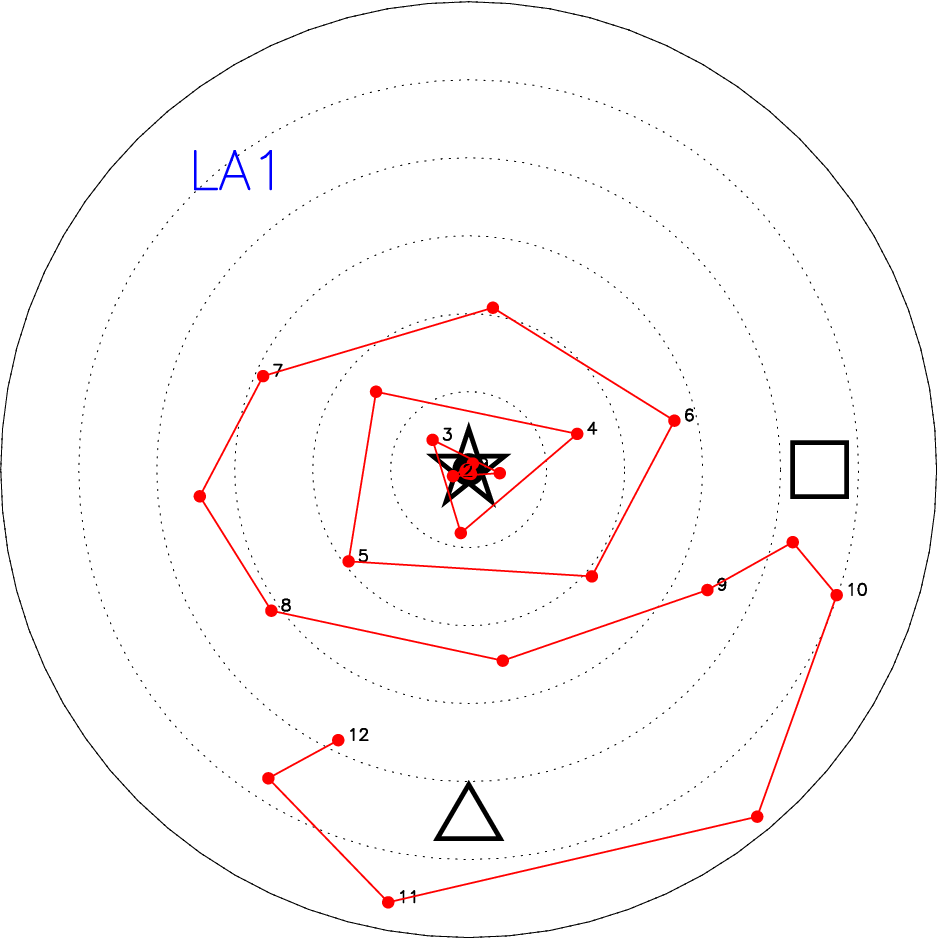}
\includegraphics[angle=0.,width=0.5\hsize]{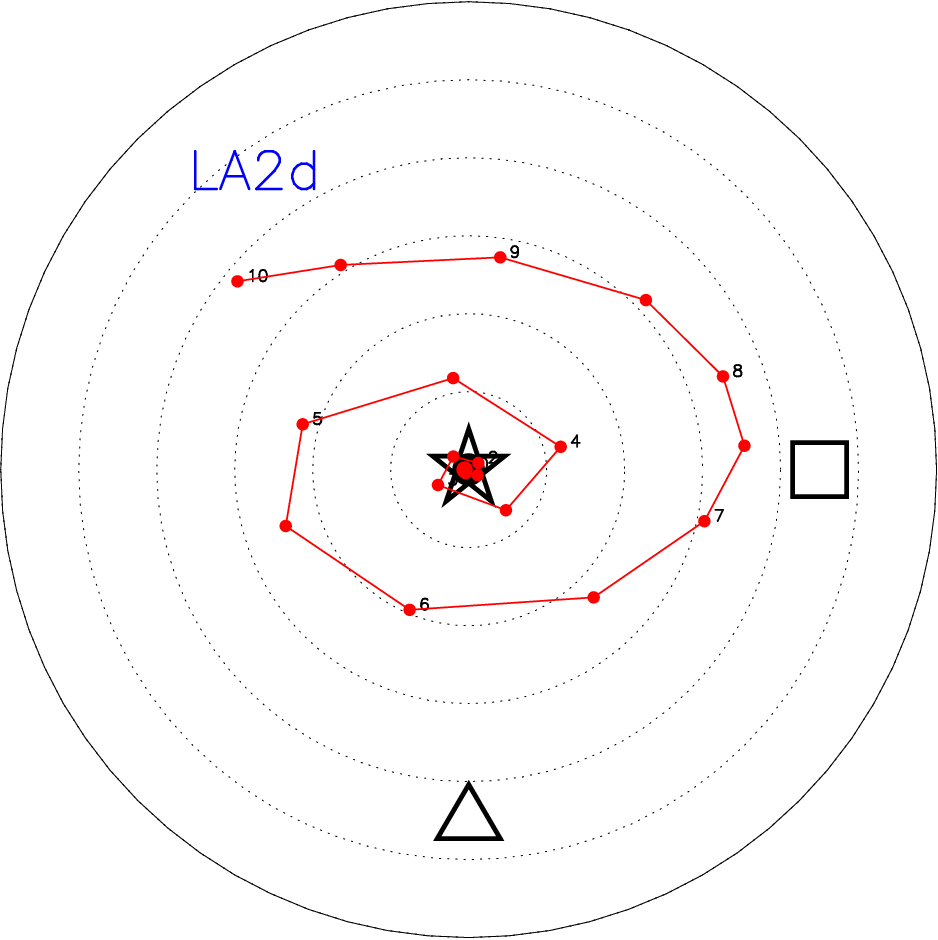}
}
\caption{ Briggs figures (see Section \ref{ssec:briggsfigs} for a
  explanation of these figures) showing the evolution of models SA1
  (top-left), IA3 (top-right), LA1 (bottom-left) and LA2d
  (bottom-right).  Dotted circles are spaced at $20\degrees$
  intervals, with the outer solid circle corresponding to $\theta =
  120\degrees$ in each case.  The disc spin is indicated by the big
  filled (black) circles at $t=0$ and by the small filled (red)
  circles for later times.  The small numbers besides some of the disc
  spins indicate the time.  The open (black) star, square and triangle
  symbols indicate the direction of the long, intermediate and short
  axes of the haloes on large scales.  
\label{fig:collisionlessbriggs}}
\end{figure*}

\subsection{Evolution of the Short-Axis Orientation}
\label{ssec:minoraxis}

The top-left panel of Fig. \ref{fig:collisionlessbriggs} shows the
evolution of model SA1 which starts with its disc spin along the short
axis of the halo.  The disc remains co-planar in this orientation for
10 \Gyr, despite its low mass.  The net potential is everywhere
flattened in the same direction as the disc, even when only the halo
potential is considered.  Small tilts of the disc therefore lead to
torques which drive damped precession but do not lead to a runaway
tilt, making this orientation stable.

\subsection{Evolution of the Intermediate-Axis Orientation}
\label{sec:intaxis}

D13 showed that discs grown in halo A at the intermediate-axis
orientation tilt very rapidly.  They showed that this occurs even when
the potential surrounding the disc is flattened by it to the extent
that its spin is along the potential's short axis.  D13 argued that
this indicates that the halo itself is unstable in this orientation.
They presented evidence that orbits of dark matter particles that
cross between the inner, flattened halo, and the outer elongated halo
are unstable, which they proposed is the cause of the halo
instability.

\begin{figure}
\includegraphics[angle=-90.,width=\hsize]{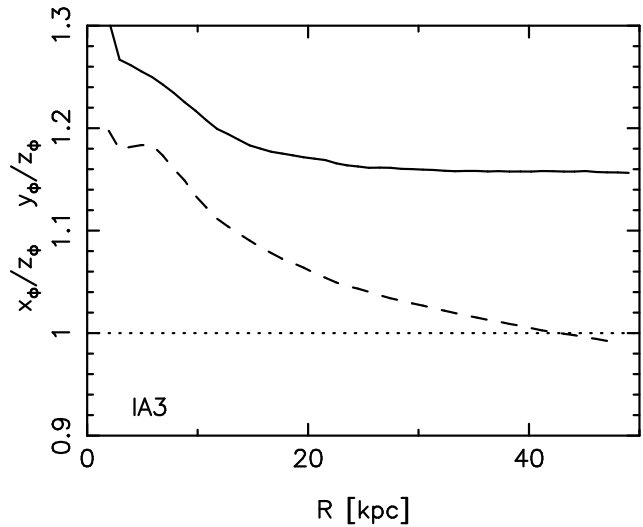}
\caption{Halo only equipotential axes ratios in model IA3.  The solid
  line shows $x_\Phi/z_\Phi$ while the dashed line shows
  $y_\Phi/z_\Phi$.  The $z$-axis is perpendicular to the disc at
  $t=0$.  
\label{fig:potshapeIA3}}
\end{figure}

Here, we present another simulation, model IA3, of a disc in halo A
with a mass $\Mdisc = 7\times10^{11} \msun$, \ie\ four times more
massive than the disc in model IA1, the most massive disc in halo A
presented by D13.  Fig. \ref{fig:potshapeIA3} shows the axes-ratios of
the halo potential along the principal axes $x$, $y$ and $z$, with the
$z$-axis perpendicular to the plane of the disc at $t=0$.  This figure
shows that $x_\Phi/z_\Phi > 1$ and $y_\Phi/z_\Phi > 1$, \ie\ the
high-mass disc has flattened the halo so much that the short axis of
the halo potential is vertical to the disc all the way out to 40 kpc.
If the disc potential is included also, then the global potential
becomes even more flattened along the disc's spin axis.  Therefore,
the disc, out to large radii, sits in a potential which never has its
intermediate axis parallel to the disc spin axis.  Naively then this
orientation may be expected to be stable.  Indeed when we evolve the
system with the halo particles frozen in place, the disc remains in
this orientation for 10 Gyr.  However, this orientation is not stable
when the halo particles are live, as can be seen in the top-right
panel of Fig. \ref{fig:collisionlessbriggs}.  As did model IA1 of D13,
the disc tilts away very rapidly from this orientation to a short-axis
orientation.

\subsection{Evolution of the Long-Axis Orientation}
\label{sec:majoraxis}

The evolution of stellar discs placed with their spin along the long
axis of the halo depends on the shape of the halo potential within
which the disc is immersed.  In model LA1, the low-mass disc is quickly
driven off its original plane and precesses rapidly around the long
axis while slowly tilting towards an orthogonal orientation.  This
precession ends abruptly when the disc reaches a nearly
intermediate-axis orientation; after this the disc drops towards a
short-axis orientation, precessing slowly about this orientation, as
shown in bottom-left panel of Fig. \ref{fig:collisionlessbriggs}.
The more massive disc in model LA2 does not tilt much, only reaching
$\theta \simeq 15\degrees$ after 10 \Gyr, but it precesses about the
long axis.

\begin{figure}
\includegraphics[angle=0.,width=\hsize]{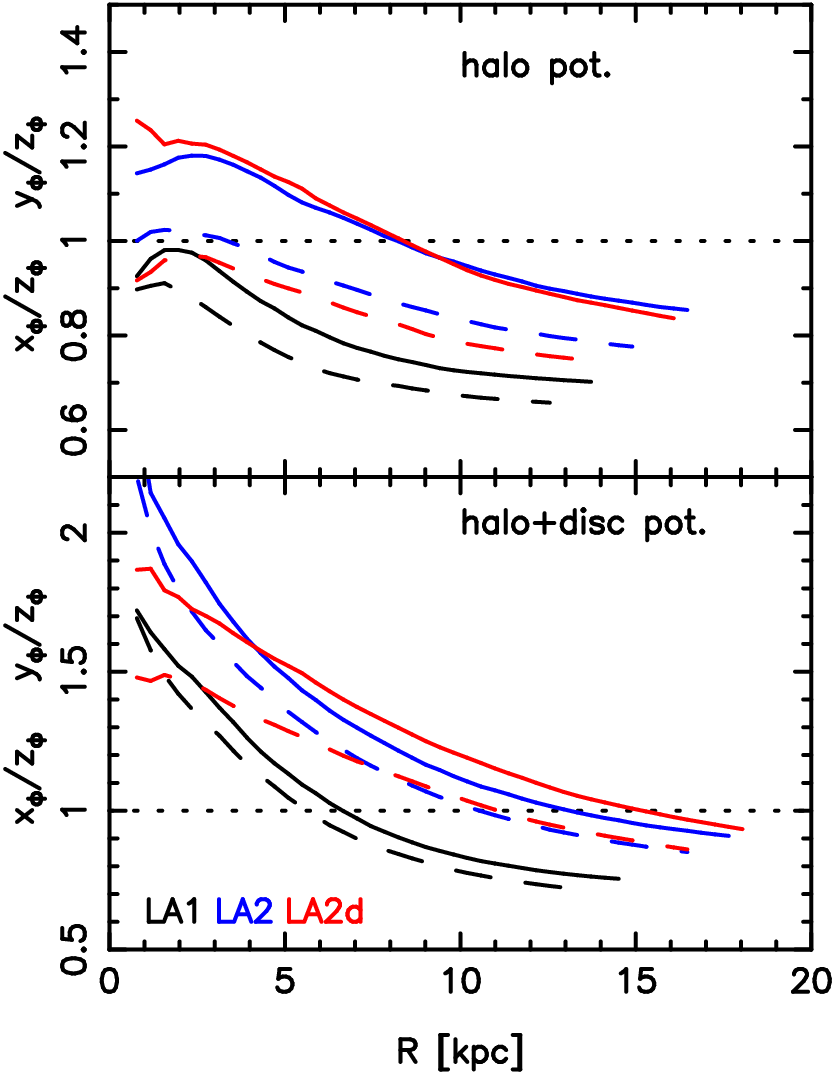}
\caption{Equipotential axes ratios in models LA1 (black lines), LA2
  (blue lines) and LA2d (red lines).  The solid lines show
  $x_\Phi/z_\Phi$ while the dashed lines show $y_\Phi/z_\Phi$, where
  $z > x > y$ are the halo's long, intermediate and short axes on
  large scales and the $z$-axis is perpendicular to the disc.  The
  dotted horizontal lines indicate an axis ratio of unity; if both
  axes-ratios follow this line then the potential is spherical.  The
  top panel shows the shape of the halo potential only, while the
  bottom one shows the shape of the full potential.
\label{fig:potshapesLA}}
\end{figure}

\begin{figure}
\centerline{
\includegraphics[angle=-90.,width=\hsize]{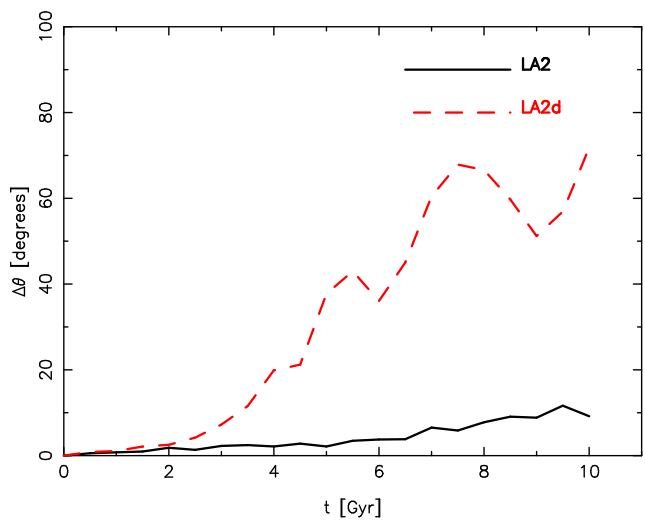}
}
\caption{ Evolution of the angle between the instantaneous stellar
  angular momentum and the initial one in models LA2 and LA2d.  Both
  models start in a long-axis orientation in the same halo and have
  the same mass.  Model LA2d has a scale-length twice that of LA2.
  \label{fig:potshapes2}}
\end{figure}

An explanation for why LA1 tilts while LA2 appears stable can be
obtained from Fig. \ref{fig:potshapesLA}, which shows the profile of
the axis-ratios of the potential along the three principal axes, $x$,
$y$ and $z$ with the $z$-axis perpendicular to the disc.  Close to the
centre of the galaxy the equipotentials are flattened like the disc
($x_\Phi/z_\Phi \simeq y_\Phi/z_\Phi > 1$), becoming almost prolate
and orthogonal to the disc further out ($x_\Phi/z_\Phi \simeq
y_\Phi/z_\Phi < 1$).  The entire disc in model LA1 sees an orthogonal
prolate halo potential; the top panel of Fig. \ref{fig:potshapesLA}
shows that, for the halo potential, $\Phi_x/\Phi_z \simeq
\Phi_x/\Phi_z < 1$.  Thus, this orientation is unstable and any small
perturbation of the orientation leads to the disc being torqued out of
its original plane.  In model LA2, instead a large part of the disc is
surrounded by a halo potential which is vertically flattened
($\Phi_x/\Phi_z > 1$).  Thus, small perturbations do not lead to
strong torques in model LA2, so it lasts in this orientation longer.
We test this interpretation in model LA2d, in which a disc with the
same mass as in LA2 but with a larger scalelength, $R_{\rm d} = 6$
kpc, is grown in halo A.  The shape of the resulting potential is
presented in Fig. \ref{fig:potshapesLA}; this is not much different
from LA2.  However, now a smaller mass fraction of the disc is covered
by a vertically flattened halo potential.  As a result, while model
LA2 only tilted by $\sim 10 \degrees$, model LA2d tilts by $\sim 80
\degrees$, as shown by the Briggs figure at bottom-right of Fig.
\ref{fig:collisionlessbriggs}.  Fig. \ref{fig:potshapes2} compares the
evolution of the tilt in LA2 and LA2d; while the tilt angle of model
LA2 grows slowly, model LA2d, which has about twice the angular
momentum, nonetheless tilts rapidly.  This shows that it is the disc
material sitting outside the vertically flattened potential that is
responsible for the fast tilting in the long-axis orientation.

\subsection{Synthesis of collisionless simulations}

\begin{figure}
\centerline{\includegraphics[angle=-90.,width=\hsize]{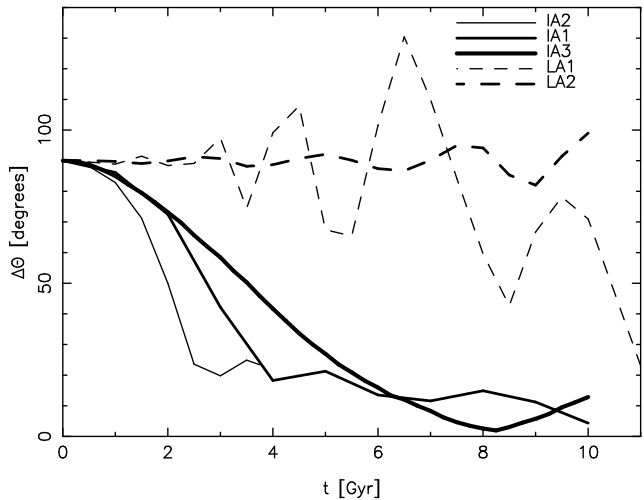}}
\caption{ Evolution of the angle between the stellar disc angular
  momentum and the halo short axis at large radius.  Solid lines are
  models with discs starting in the intermediate-axis orientation and
  dashed lines are for those starting in the long-axis orientation, as
  indicated at top-right.  All models are evolved in halo A.
\label{fig:angmomang}}
\end{figure}

Fig. \ref{fig:angmomang} compares the tilting of all the long- and
intermediate-axis simulations in halo A.  The intermediate-axis
simulations most rapidly reach the short-axis orientation.  At low
disc mass, the long-axis orientation is unstable with the disc tilting
and precessing.  At high disc mass, the long-axis orientation is still
unstable, but now the tilting rate is so low that the disc persists
close to this orientation for most of a Hubble time.  However, the
fact that this orientation is only quasi-stable suggests that it may
be susceptible to perturbations from satellites, as we demonstrate
below.  

Comparing the discs during the period when the disc mass is growing
(with the disc particles fixed in place and the halo particles free to
move), we find that the disc in the short-axis orientation has a
potential energy $\sim 1\%$ smaller (more bound) than the disc in the
long-axis orientation for the same mass, making it the most stable
orientation.  In a fixed halo potential, it is readily apparent that
the disc potential is minimized when the disc is in a short-axis
orientation; this is still true even when the inner halo is flattened
by the disc growing within it.

\section{The Impact of Cooling Gas}
\label{sec:gasruns}

Although dark matter angular momentum tends to align with the halo's
short axis, cosmological simulations show that the gas angular
momentum is frequently decoupled from that of the halo
\citep{vdb_etal_02, che_etal_03, bailin+05, sha_ste_05, roskar+10}.
We now explore how the delivery of gas with angular momentum along
different directions affects the evolution of disc orientation.

\subsection{Initial conditions with gas}
\label{ssec:gasics}

\begin{figure}
\includegraphics[angle=0.,width=\hsize]{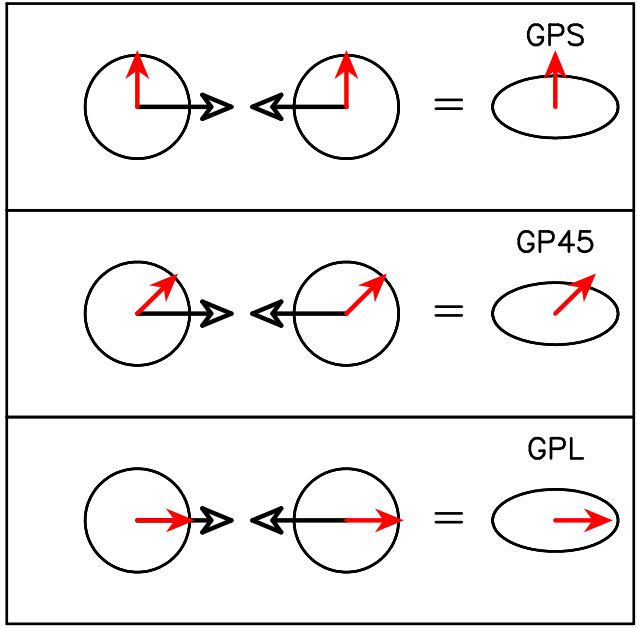}
\caption{Cartoons showing the merger geometries that produce the
  prolate models GPS, GP45 and GPL.  The (red) filled arrows indicate
  the orientation of the gas angular momentum, while the (black) open
  arrows show the merging velocities of the spherical haloes.  The
  final long axis of the merger remnant is along the merging
  direction.}
\label{fig:mergergeometryprolates}
\end{figure}

\begin{figure}
\centerline{\includegraphics[angle=-90.,width=\hsize]{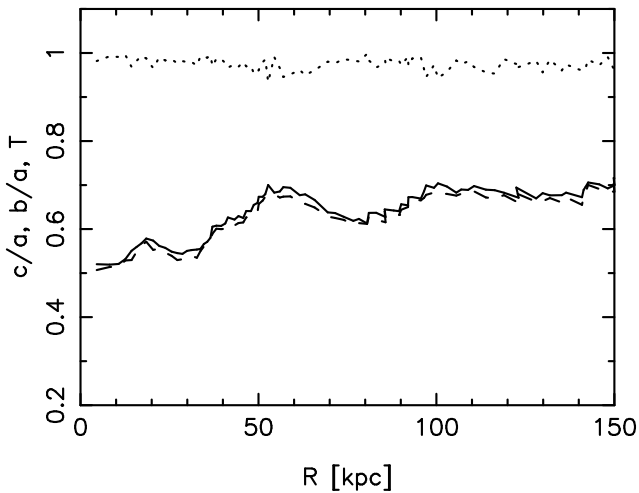}}
\caption{The shape of the dark matter density for the haloes of models
  GPL, GPS and GP45.  Solid, dashed and dotted lines show $b/a$, $c/a$
  and $T$, respectively.  The three haloes have identical shape.
\label{fig:gashaloshapes}}
\end{figure}

We perform experiments of discs forming out of gas cooling within both
prolate and fully triaxial haloes.  All stars in these simulations
form self-consistently out of the cooling gas.  As did
\citet{aumer_white13}, in early experiments we found that arbitrarily
inserting rotating gas coronae within pre-existing triaxial dark
matter haloes leads to a substantial loss of gas angular momentum,
making the formation of realistic disc galaxies difficult.  Our
approach therefore is to include the gas right from the start while
constructing the prolate/triaxial dark matter haloes via mergers.  We
set up a prolate halo with an equilibrium gas distribution by the
head-on merger of two spherical Navarro-Frenk-White dark matter haloes
each with an embedded spherical hot gas component containing $10\%$ of
the total mass and following the same density distribution.  The
starting haloes are set up as described in \citet{roskar+08a}: each
has a mass within the virial radius ($r_{200} \simeq 200 \kpc$) of
$10^{12} \msun$.  A temperature gradient in each corona ensures an
initial gas pressure equilibrium for an adiabatic equation of state.
Gas velocities are initialized to give a spin parameter of $\lambda =
0.065$ \citep{bullock_angmom+01, maccio+07b}, with specific angular
momentum $j \propto R$, where $R$ is the cylindrical radius.  Each
halo uses $1\times10^6$ particles in each of the gas and dark
components.  Gas particles initially have masses $1.4\times 10^5
\msun$ and softening 50 pc, the latter inherited by the star
particles, while dark matter particles come in two mass flavors
($10^6\msun$ and $3.5\times10^6 \msun$ inside and outside 200 kpc,
respectively) and with a softening of 100~pc.  Initially the two
haloes are 500 kpc apart and moving towards each other at a relative
velocity of 100~\kms, producing prolate systems.  We generate three
such prolate haloes differing by the orientation of the gas angular
momentum relative to the halo long axis.  Since the remnant's long
axis is along the initial separation vector, we only need to incline
the initial spherical gas haloes relative to the separation vector to
ensure a remnant with gas angular momentum tilted relative to the halo
long axis.  We chose tilt angles of $0\degrees$, $45\degrees$ and
$90\degrees$ (models GPL, GP45, and GPS, respectively).
Fig. \ref{fig:mergergeometryprolates} shows the merger geometries of
the three prolate haloes.  The final haloes are quite prolate, $T \ga
0.95$, and have $\left<c/a\right> \simeq 0.65$ (see
Fig. \ref{fig:gashaloshapes}).

\begin{figure}
\includegraphics[angle=0.,width=\hsize]{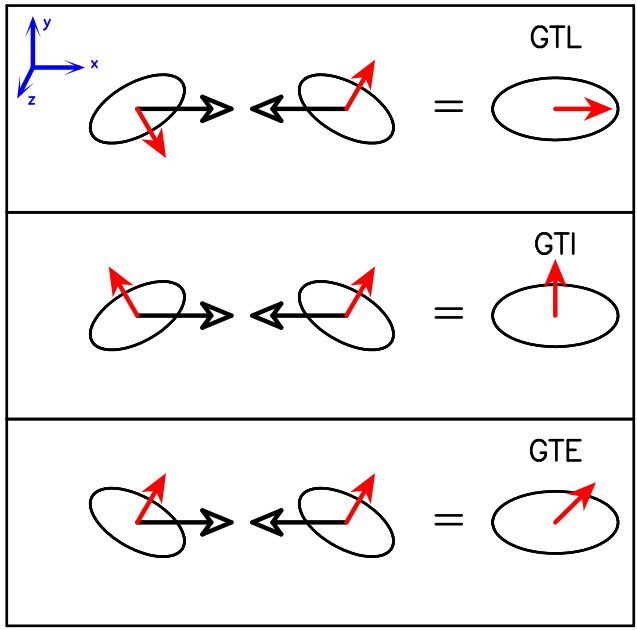}
\caption{Cartoons showing the merger geometries that produce the
  triaxial models GTL, GTI and GTE.  The (red) filled arrows indicate
  the orientation of the gas angular momentum relative to the haloes,
  while the (black) open arrows show the merging velocities of the
  spherical haloes.  The coordinate axes are indicated in the top
  panel: the long axis of the merger remnant is along the $x$-axis
  while the short axis is perpendicular to the sheet ($z$-axis).  The
  intermediate axis is along the $y$-axis.}
\label{fig:mergergeometrytriaxial}
\end{figure}

We produce triaxial models by merging two copies of the prolate system
GPS.  By judicious orientation of the angular momenta before the
merger, we are able to produce systems with angular momentum along
either the long or the intermediate axes, which we refer to as models
GTL and GTI, respectively.  Model GTI was presented in D13.  In models
GTI and GTL, we first rotate GPS about the long axis so the angular
momentum vector is along the $y$-axis, then rotate the system about
the $z$-axis by $\pm 30\degrees$.  Merging these from a separation of
500 kpc with a relative velocity of 100~\kms\ produces a triaxial
system with long, intermediate and short axes corresponding to the
$x$, $y$ and $z$ axes, respectively.  The merger geometries for these
models are shown in Fig. \ref{fig:mergergeometrytriaxial}.  We also
ran an additional model with gas angular momentum initially in the
plane spanned by the long and intermediate axes (model GTE).  We
produce this model by rotating GPS about the $y$-axis by $90\degrees$
as before, then merging two copies of this system {\it both} rotated
about the $z$-axis by $-30\degrees$.  This merger geometry is also
shown in Fig. \ref{fig:mergergeometrytriaxial}.  By construction, none
of the prolate and triaxial haloes have any figure rotation.  The
shape of the three triaxial haloes is identical and is shown in Figure
1 of D13.

These simulations are evolved with {\sc gasoline} \citep{gasoline},
the smoothed particle hydrodynamics (SPH) version of {\sc pkdgrav}.
We use a timestep of 10 Myr with a refinement parameter $\eta =
0.175$.  During the merger, and for some time thereafter, we evolve
the system adiabatically without any star formation.  Then we switch
on gas cooling, star formation and stellar feedback using the
prescriptions of \citet{stinson+06}.
A gas particle undergoes star formation if it has number density $n >
0.1 {\mathrm cm}^{-3}$, temperature $T < 15,000$ K and is part of a
converging flow; efficiency of star formation is 0.05, \ie\ 5\% of gas
eligible to form stars does so per dynamical time.  Star particles
form with an initial mass of $1/3$ that of the gas particle, which at
our resolution corresponds to $4.6 \times 10^4 \msun$.  Gas particles
can spawn multiple star particles but once they drop below $1/5$ of
their initial mass the remaining mass is distributed amongst the
nearest neighbours, leading to a decreasing number of gas particles.
Each star particle represents an entire stellar population with a
Miller-Scalo \citep{miller_scalo79} initial mass function. The
evolution of star particles includes feedback from Type II and Type Ia
supernovae, with their energy injected into the interstellar medium
(ISM). The effect of the supernovae explosions is mode,led at the
sub-grid level as a blastwave propagating through the ISM
\citep{stinson+06}. We also include feedback from asymptotic giant
branch stellar winds.
We use an opening angle of $\theta = 0.7$.  The timestep of gas
particles also satisfies the condition $\delta t_{gas} =
\eta_{courant} h/[(1+\alpha)c + \beta\mu_{max}]$, where
$\eta_{courant} = 0.4$, $h$ is the SPH smoothing length, $\alpha$ is
the shear coefficient, which is set to 1, $\beta=2$ is the viscosity
coefficient and $\mu_{max}$ is described in \citet{gasoline}.  The SPH
kernel is defined using the 32 nearest neighbours.  Gas cooling is
calculated without taking into account the gas metallicity.  These
prescriptions have been shown to lead to realistic Milky Way-type
galaxies \citep{roskar+12, roskar+13}.  We set $t=0$ to be the time at
which we switch on gas cooling and star formation.

\begin{centering}
\begin{table}
\vbox{\hfil
\begin{tabular}{lcccc}\hline 
\multicolumn{1}{c}{Model} &
\multicolumn{1}{c}{Halo} &
\multicolumn{1}{c}{$N_*$} &
\multicolumn{1}{c}{$M_*$} &
\multicolumn{1}{c}{$\lambda$} \\

 & & $(10^6)$ & ($10^{10} \msun$) & \\ \hline

%
GPS & GP & 3.16 & 9.4 & 0.058 \\
%
GP45& GP & 3.16 & 9.5 & 0.055 \\
%
GPL & GP & 3.29 & 9.9 & 0.047 \\
%
GTI & GT & 6.97 & 20.7 & 0.021 \\
%
GTL & GT & 7.10 & 21.1 & 0.020 \\
%
GTE & GT & 6.69 & 19.9 & 0.029 \\ \hline

\end{tabular}
\hfil}
\caption{ The simulations with gas and star formation.  $N_*$ and
  $M_*$ are the number and the mass of star particles at the final
  step.  $\lambda$ is the angular momentum parameter (computed using
  equation \ref{eqn:lambda}) at $t=0$, when we turn on gas cooling and
  star formation.  }
\label{tab:gassims}
\end{table}
\end{centering}

Table \ref{tab:gassims} lists all the gas$+$star formation
simulations.  It includes the value of $\lambda$
\citep{bullock_angmom+01}, the angular momentum parameter of the gas.
This is defined as
\begin{equation}
\lambda = \frac{L_g}{G M_t}\sqrt{\frac{|E_g|}{M_g^3}}
\label{eqn:lambda}
\end{equation}
where $L_g$, $E_g$ and $M_g$ is the angular momentum, energy and mass
of the gas within the virial radius, and $M_t$ is the total
(gas$+$dark matter) mass within the same radius.  As Fig.
\ref{fig:mergergeometryprolates} suggests, the merger geometry should
result in all the prolate halo models having the same gas angular
momentum.  Instead $\lambda$ increases as the angular momentum
direction changes from the long to the short axis, indicating that
angular momentum is being transported within the gas corona.

\subsection{Evolution of disc orientation in prolate haloes with gas}


We start by considering the simpler evolution of discs forming in
prolate haloes in which the initial gas angular momentum is inclined to
the halo long axis by $0\degrees$ (model GPL), $45\degrees$ (model
GP45) and $90\degrees$ (model GPS).

After 8 \Gyr, all three systems form stellar discs of comparable
mass.  Model GPS forms a stellar disc with its spin perpendicular to
the halo's long axis and remains in this orientation throughout its
evolution.  Within 1 Gyr the disc spin in model GPL settles almost
into alignment with the long axis to within $5\degrees$, where it
remains.  Even when we shut off gas cooling at 4 Gyr, the disc still
remains in a long-axis orientation, in good agreement with the results
of the collisionless simulations.

\begin{figure*}
\centerline{
\includegraphics[angle=0.,width=0.5\hsize]{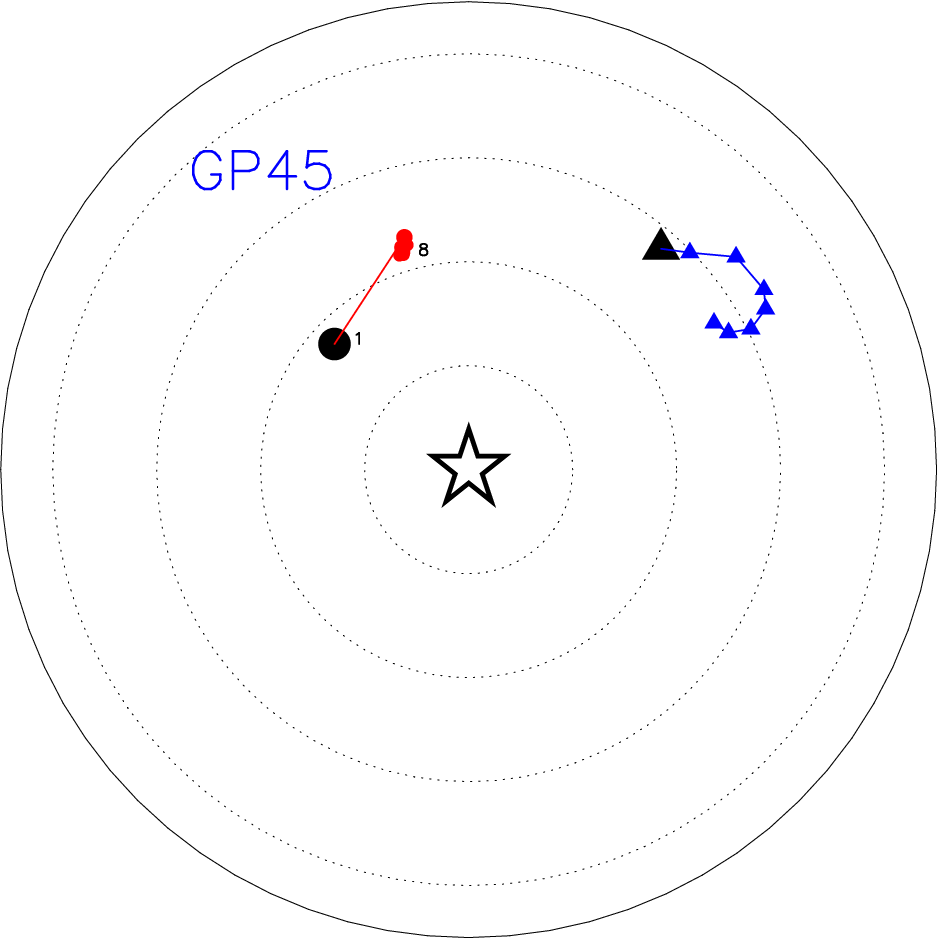}
\includegraphics[angle=0.,width=0.5\hsize]{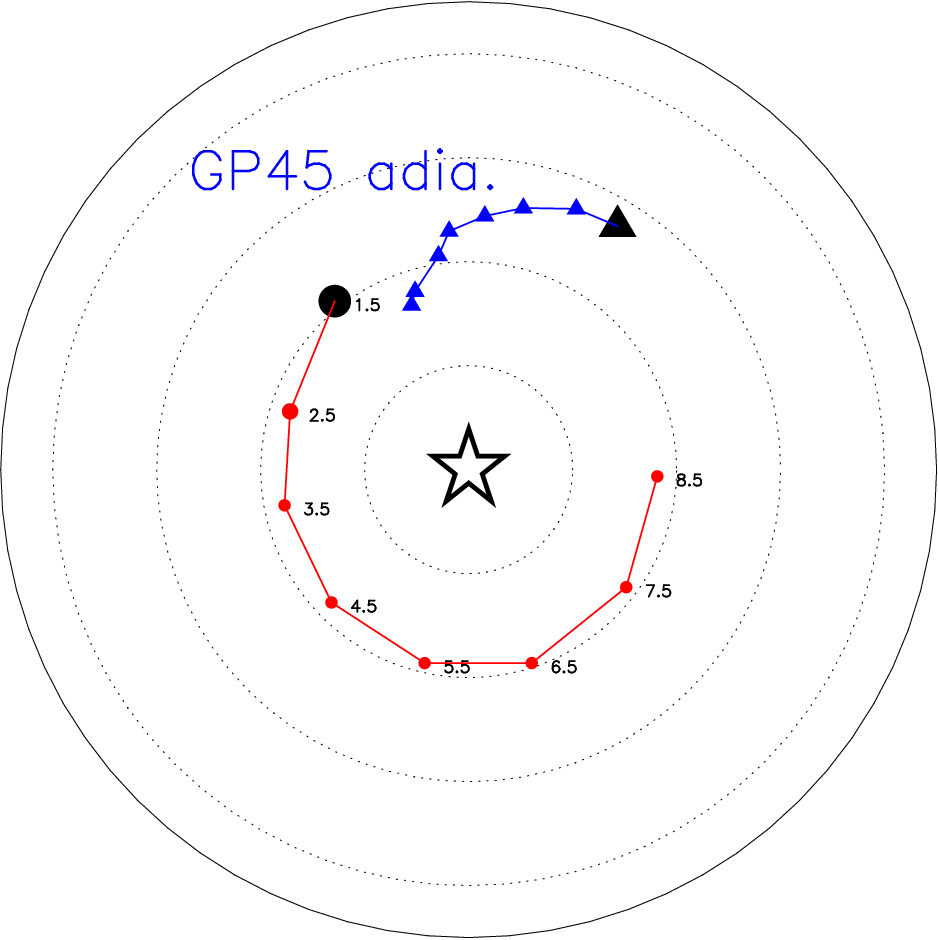}
}
\caption{Left: Briggs figure for model GP45 showing 8 \Gyr\ of
  evolution.  Right: Briggs figure for model GP45 when gas cooling is
  turned off after 1 \Gyr.  Dotted circles are spaced by $20\degrees$.
  The disc spin is indicated by the filled (red and black) circles.
  The (blue and black) filled triangles mark the orientation of the
  hot gas angular momentum between 20 \kpc\ and 50 \kpc.  The spin of
  both the disc and gas at 1 \Gyr\ (left) and 1.5 \Gyr\ (right) are
  marked by the larger (black) corresponding symbols.  The long axis
  corresponds to the open (black) star.  The outer solid circle
  corresponds to $\theta = 90\degrees$; since this is a prolate
  (axisymmetric) halo each point on this circle corresponds to a short
  axis of the halo.  
\label{fig:G45briggs}}
\end{figure*}

\begin{figure*}
\centerline{
\includegraphics[angle=0.,width=0.5\hsize]{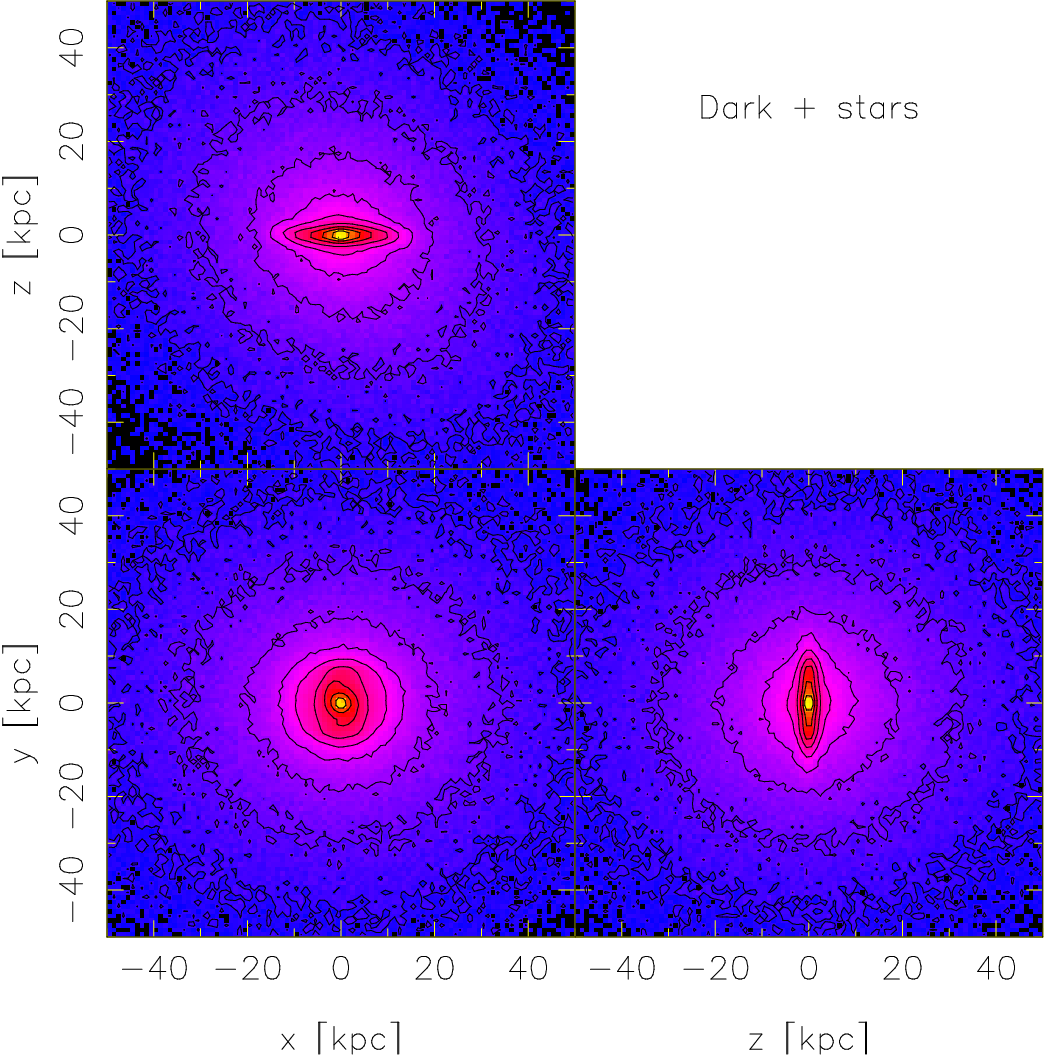}
\includegraphics[angle=0.,width=0.5\hsize]{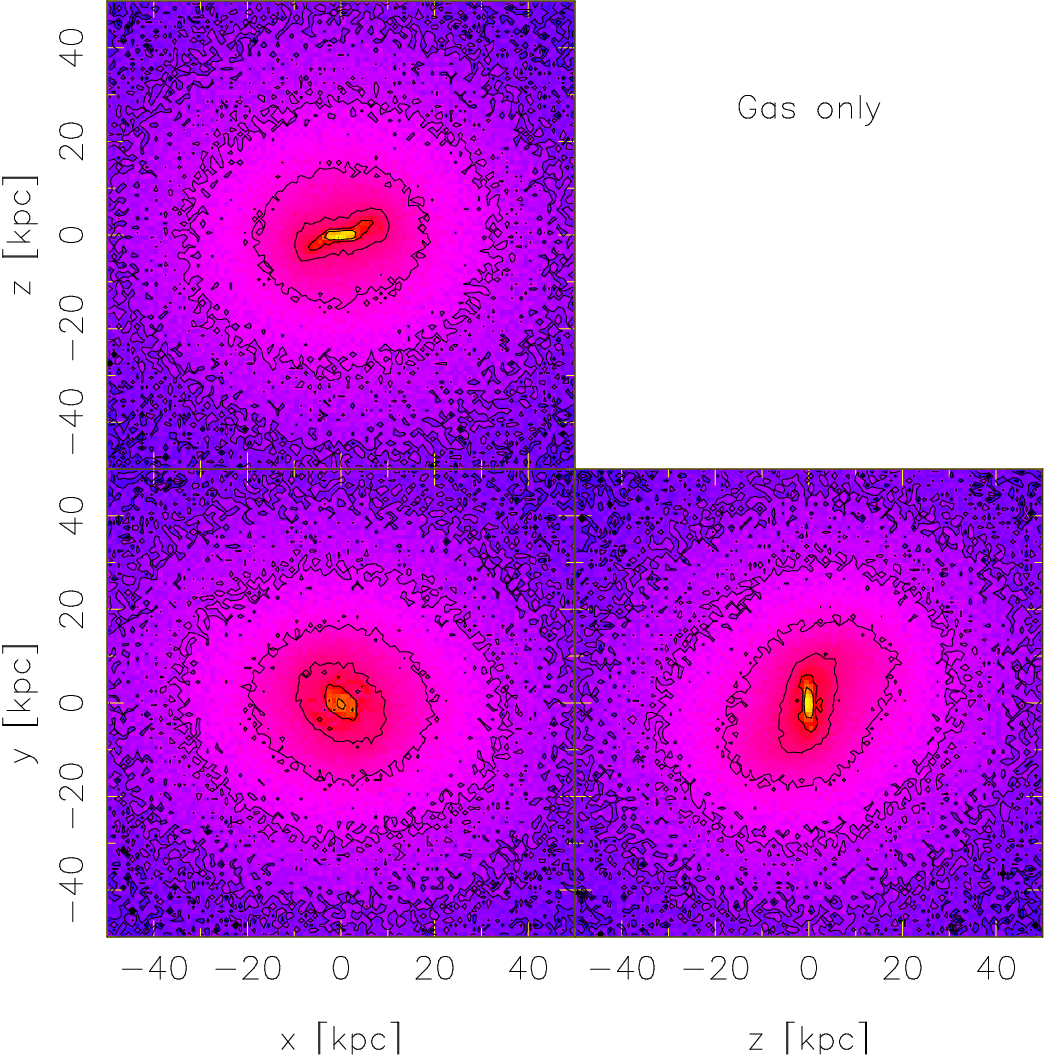}}
\caption{Model GP45 at 8 Gyr with the system rotated so the disc is in
  the $(x,y)$-plane and the halo long-axis is in the $(x,z)$-plane.
  Left: Disc$+$halo mass distribution.  The contribution of the stars
  has been increased by a factor of $10\times$ to make the disc more
  apparent.  Right: gas mass distribution at the same orientation.
  \label{fig:G45prettypic}}
\end{figure*}

The most interesting result of these simulations comes from model
GP45.  In this model, the stellar disc never settles to one of the
main planes of the halo.  Nor does the disc line up with its angular
momentum parallel to that of the gas.  Instead, as Fig.
\ref{fig:G45briggs} shows, the two angular momenta are misaligned by
$\sim 65\degrees$ and remain in this orientation without precessing
throughout the simulation.  Fig. \ref{fig:G45prettypic} shows the
dark$+$stellar mass distribution at 8 Gyr, clearly showing a stellar
disc tilted relative to the main planes of the halo.  The halo is
misaligned with the disc already at 20 kpc, which is well within the
region where the halo of the Milky Way is probed by the Sagittarius
tidal stream.  Modelling the Sagittarius stream therefore requires
that the tilt of the dark matter halo relative to the disc also be
taken into account. Interior to this region, the halo is flattened by
the growing stellar disc.

\begin{figure}
\centerline{\includegraphics[angle=0.,width=\hsize]{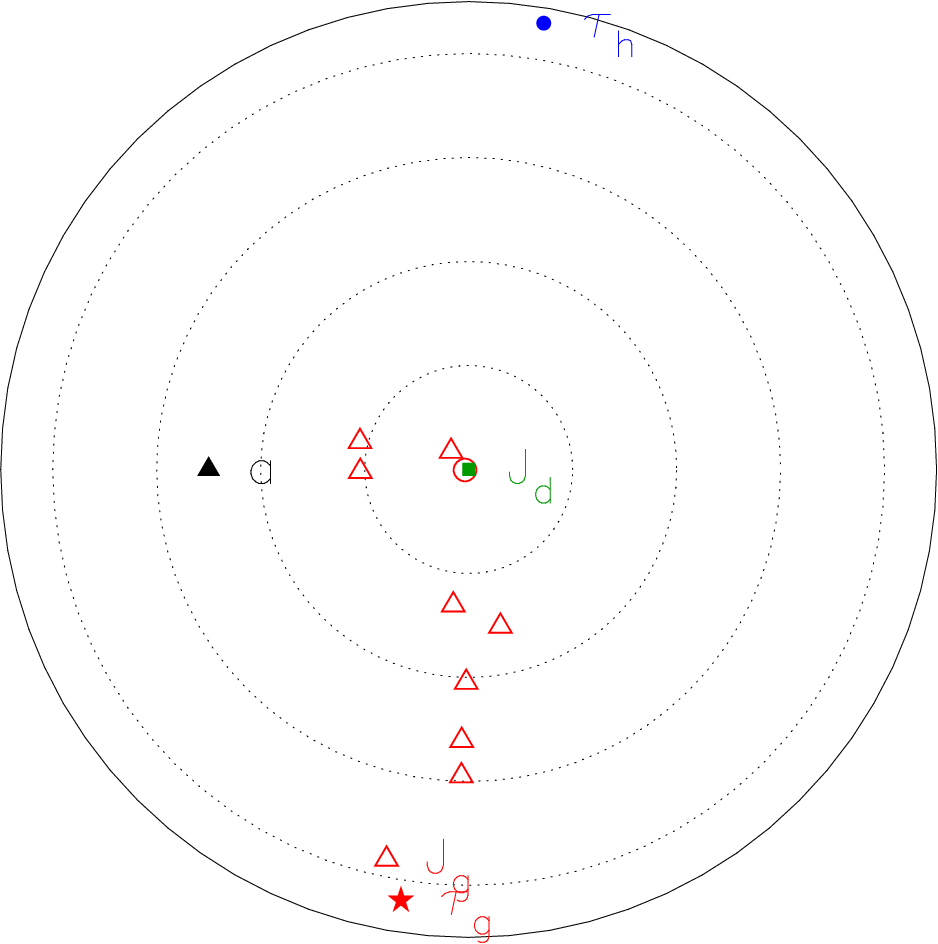}}
\caption{ Briggs figure showing the gravitational torques and angular
  momenta in model GP45 at 8 Gyr.  The dotted circles are spaced by
  $20\degrees$ intervals with the outer solid circle at $90\degrees$.
  The system has been re-oriented, as in Fig.  \ref{fig:G45prettypic},
  such that the disc's angular momentum, indicated by the solid
  (green) square, is along the $z$-axis (the origin), while the halo
  long axis (indicated by the filled, black triangle) is in the
  $(x,z)$-plane.  The solid (blue) circle indicates the torque,
  $\tau_h$, on the stars from just the dark matter halo (outside 25
  kpc) while the filled (red) star indicates the torque, $\tau_g$, on
  the stars from the gas.  The open (red) circle and triangles
  represent the {\it angular momentum} of cool ($T < 10^5$ K) gas
  within the inner 25 kpc in 10 annuli, with the central annulus
  indicated by the open circle.
\label{fig:GP45polarplot}}
\end{figure}

How is the disc able to remain in this orientation?  We compute the
torque on the stellar disc by first computing the forces on the stars
using {\sc gasoline} in the usual way.  The net torque on the disc is
then given by
\begin{equation}
{\bf{\tau}} = \sum_i m_i {\bf{r}}_i \times {\bf{F}}_i
\end{equation}
where the sum is over all star particles and $m_i$, ${\bf{r}}_i$ and
${\bf{F}}_i$ are the mass, position vector (relative to the centre of
the galaxy) and force on the i$^{th}$ star particle.  We compute the
torques on the stars from the dark matter halo ($\tau_h$) and from the
gas ($\tau_g$) separately.  Fig.  \ref{fig:GP45polarplot} presents the
torques on the stellar disc at $t=8 \Gyr$.  We also show the angular
momentum of gas within 25 kpc at $T < 10^6$~K ($J_g$); this gas, which
is destined to cool on to the stellar disc, arrives tilted with
respect to its plane.  As the gas cools from large distances, it feels
the gravitational torque from the halo and its gains angular momentum
along this direction.  Once the gas reaches the region dominated by
the stellar disc, its spin axis gains a component orthogonal to the
$\tau_g$ direction, due to the torque it feels from the disc, before
settling into the disc.  The right panel of Fig.
\ref{fig:G45prettypic} shows that the cooling gas reaches the centre
in a warped disc.  The stellar disc, therefore, is absorbing from the
cooling gas angular momentum in the direction opposite to the
direction in which the halo torque is pulling it.  As a result,
equilibrium between the gravitational torques and the gas inflow has
been established.  The equilibrium is not driven by the regulation of
the gas inflow rate since the mass of the final disc is very similar
in this model and models GPL and GPS.  Rather, the equilibrium is
determined by the disc settling into an orientation where the tilting
that would be induced by all torques on it and the tilting that the
accreting gas angular momentum would cause cancel each other out.  If
the angular momentum orientation of the cooling gas changes with time,
then the equilibrium orientation of the disc must also evolve.  We
show an example of such evolution below.

\begin{figure}
\includegraphics[angle=-90.,width=\hsize]{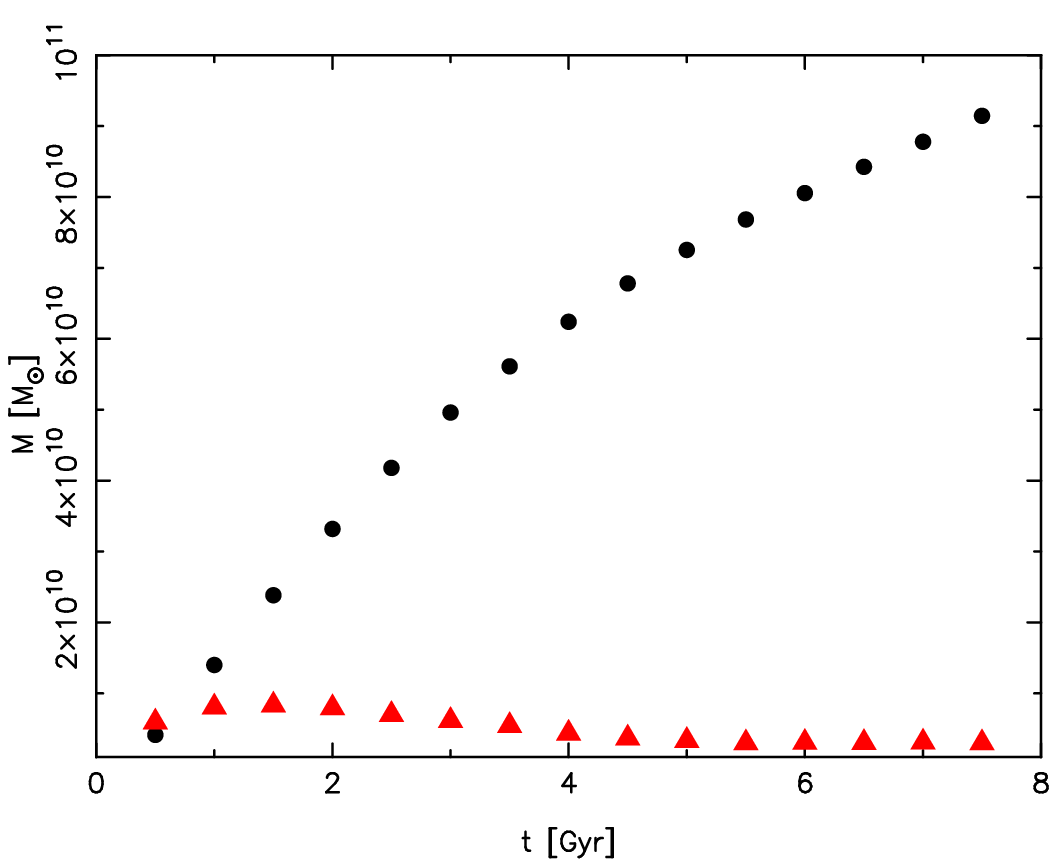}
\includegraphics[angle=-90.,width=\hsize]{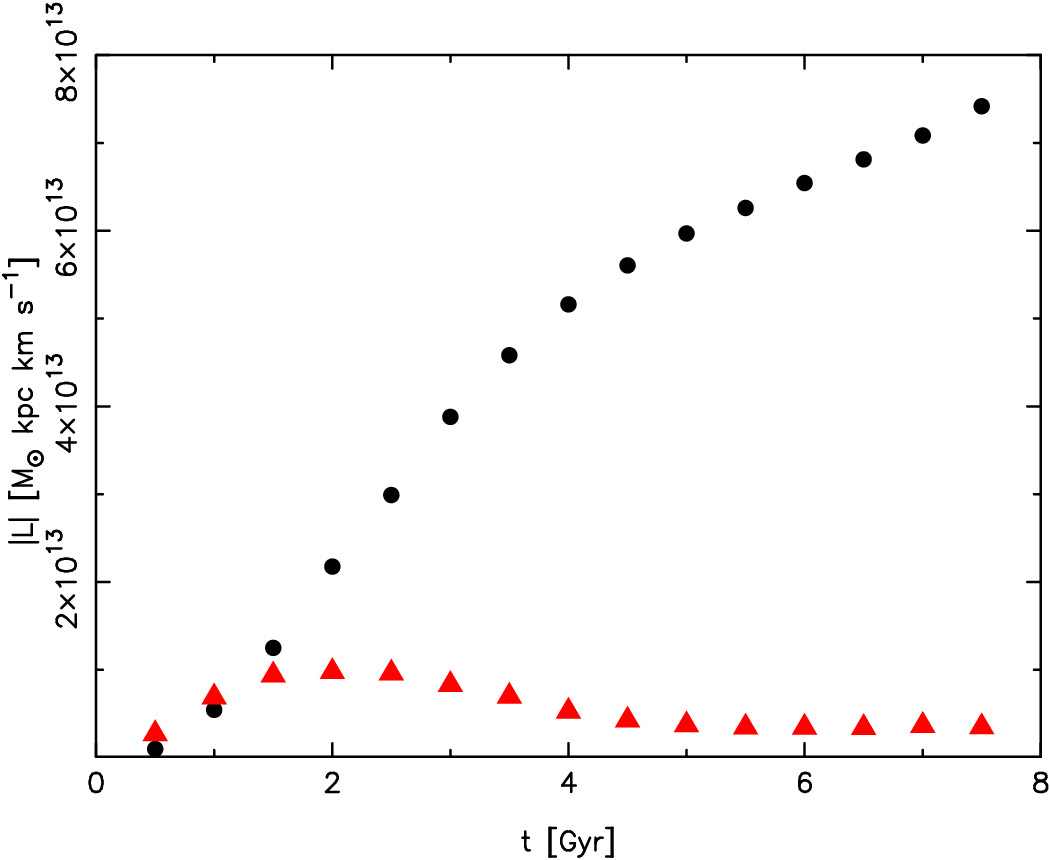}
\includegraphics[angle=-90.,width=\hsize]{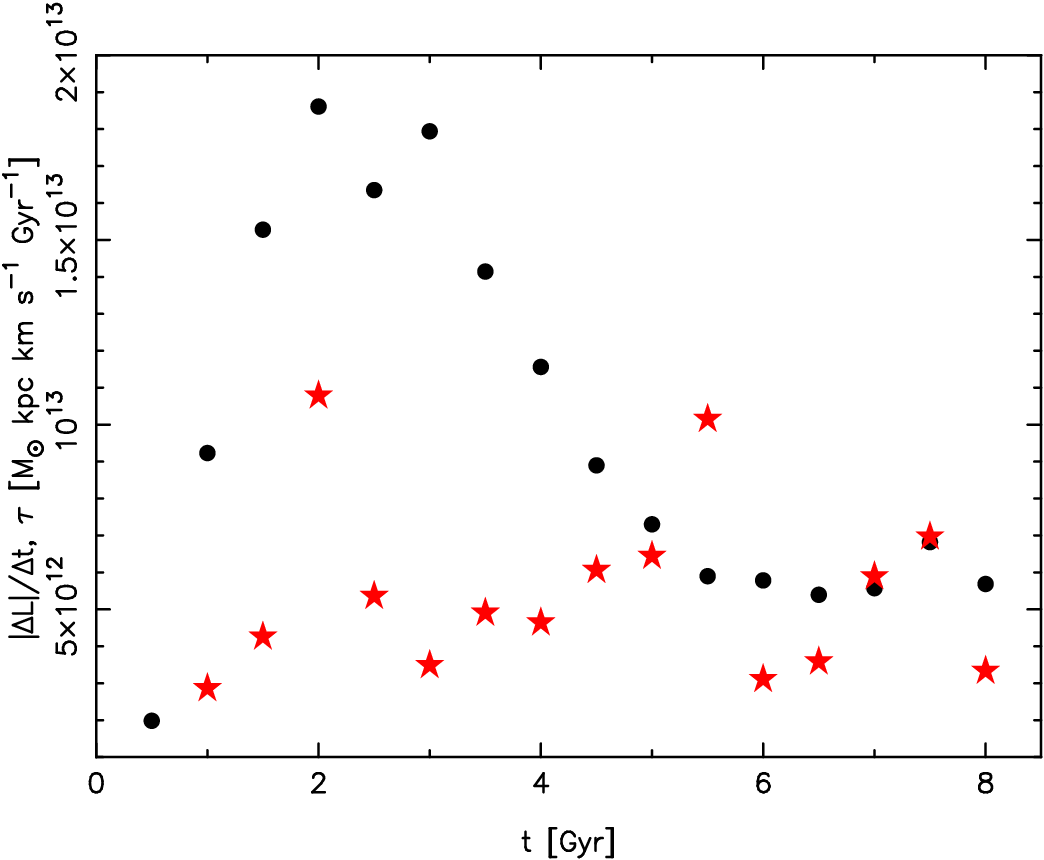}
\caption{ The evolution of the mass (top panel) and angular momentum
  (middle panel) in model GP45.  The (black) circles show the stellar
  disc while the (red) triangles show the cool ($T < 10^5$ K) gas
  disc.  The bottom panel shows the rate of change of total angular
  momentum of the stars, indicated by the (black) circles, and the
  (full) gravitational torques on the stars, indicated by the (red)
  filled stars.  
\label{fig:662momenta}}
\end{figure}

Fig. \ref{fig:662momenta} shows the evolution of the mass and angular
momentum of the stellar and cool gas discs (defined as all gas within
50 kpc at temperature $T < 10^5$ K).  By 5 Gyr the cold gas fraction
of the galaxy has fallen to $\la 5\%$ of the stellar mass.  The
stellar disc mass and angular momentum is growing at a much higher
rate than the gas disc is declining, which is possible because gas is
continuously cooling on to the disc.  Over the duration of the
simulation the cool gas net mass and angular momentum change is a
small fraction of that of the stars, indicating that the gas mass
reaches a detailed balance between inflow and loss from star formation
and feedback.  The bottom panel of Fig. \ref{fig:662momenta} compares
the gravitational torques on the stars and the rate of change of
stellar angular momentum:
\begin{equation}
  \left. \frac{|\Delta \bf{L}|}{\Delta t}\right|_{t_2} = \frac{|\bf{L}_2 - \bf{L}_1|}{t_2-t_1}.
\end{equation}
The rate of change of stellar angular momentum generally exceeds the
direct gravitational torque from the dark halo and gas corona in spite
of the strongly prolate ($b/a \la 0.7$ within 50 kpc) halo.  The
continuous delivery of angular momentum misaligned with the stellar
disc's therefore explains why the stellar disc is able to persist
misaligned relative to the halo.

When the gas can cool, the disc reaches a dynamical equilibrium and
does not change its orientation significantly; without cooling gas the
disc must precess about the long axis. An example of this precession,
with cooling turned off and evolved adiabatically after 1 Gyr (when
the stellar mass is $15\%$ of its final mass) is shown in the
right-hand panel of Fig. \ref{fig:G45briggs}.  The precession is not
accompanied by a significant tilting away from the long axis in this
prolate halo.  We also verified that the disc precesses without
tilting significantly when gas cooling is shut off at a later time
when the stellar disc is more massive.

\subsection{Evolution of disc orientation in triaxial haloes with gas}


Having understood the role of gas cooling in establishing a dynamical
equilibrium which maintains stellar discs off the principal planes in
prolate haloes, we now consider the more complicated fully triaxial
models.

\begin{figure*}
\centerline{
\includegraphics[angle=0.,width=0.5\hsize]{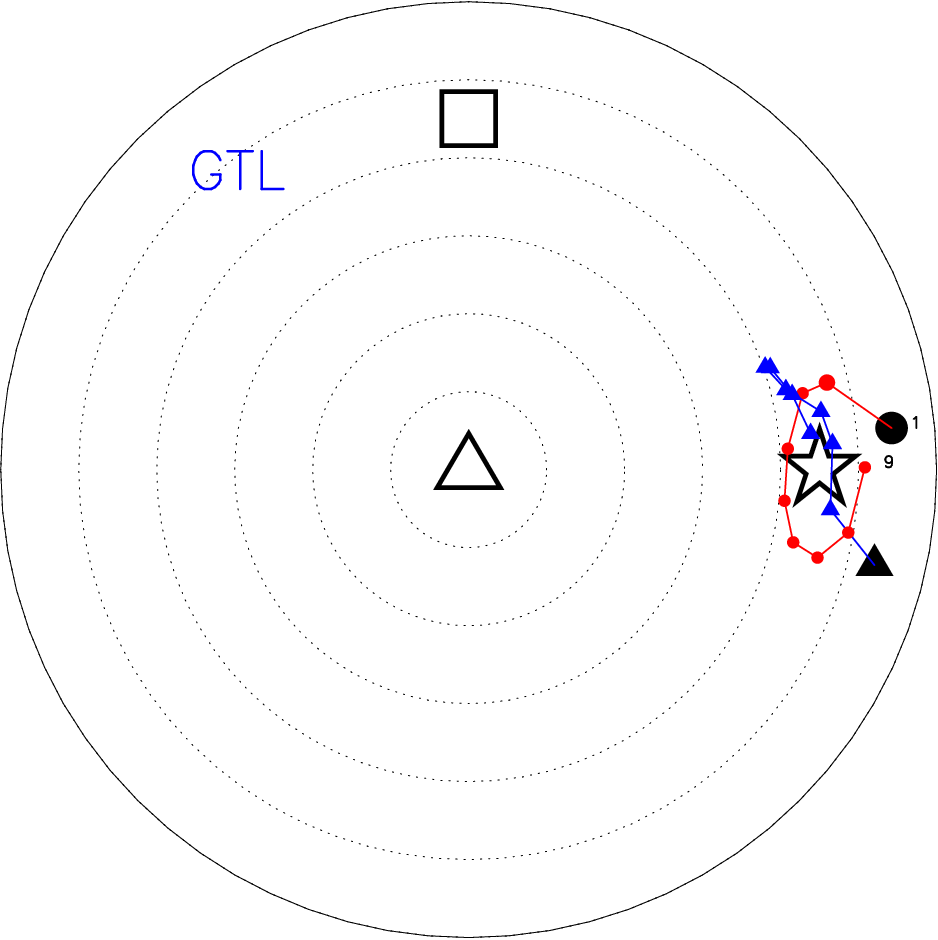}
\includegraphics[angle=0.,width=0.5\hsize]{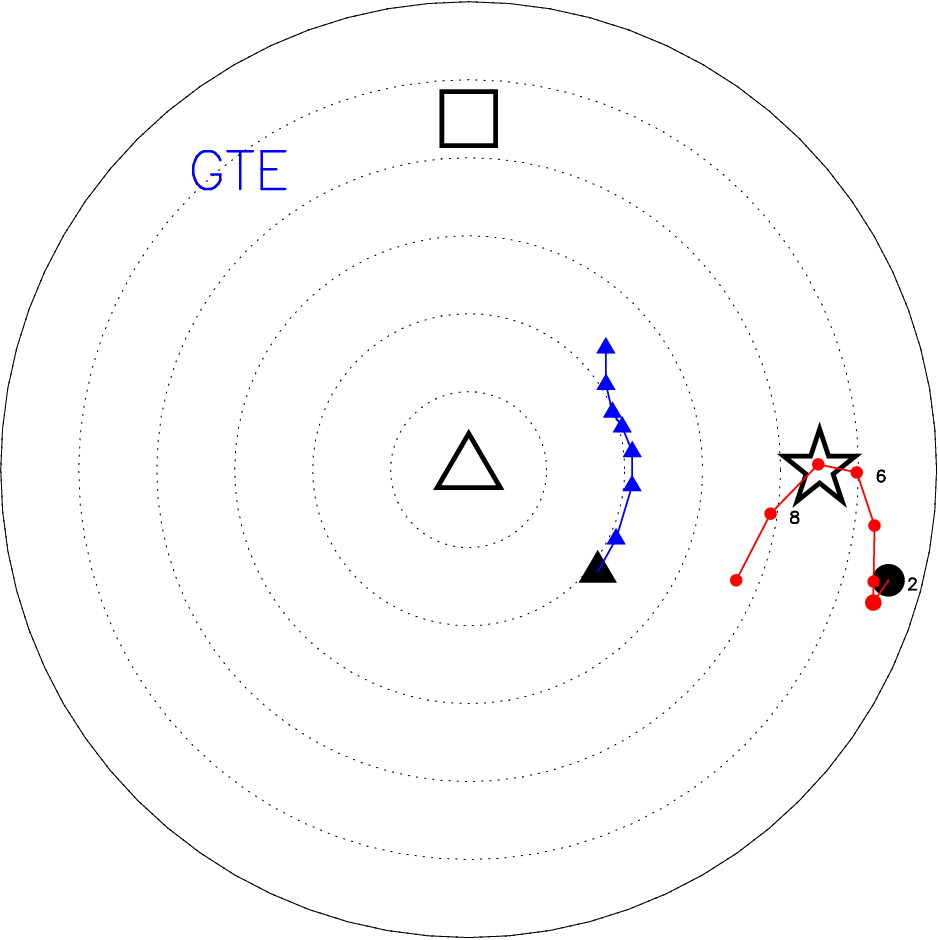}
}
\centerline{
\includegraphics[angle=0.,width=0.5\hsize]{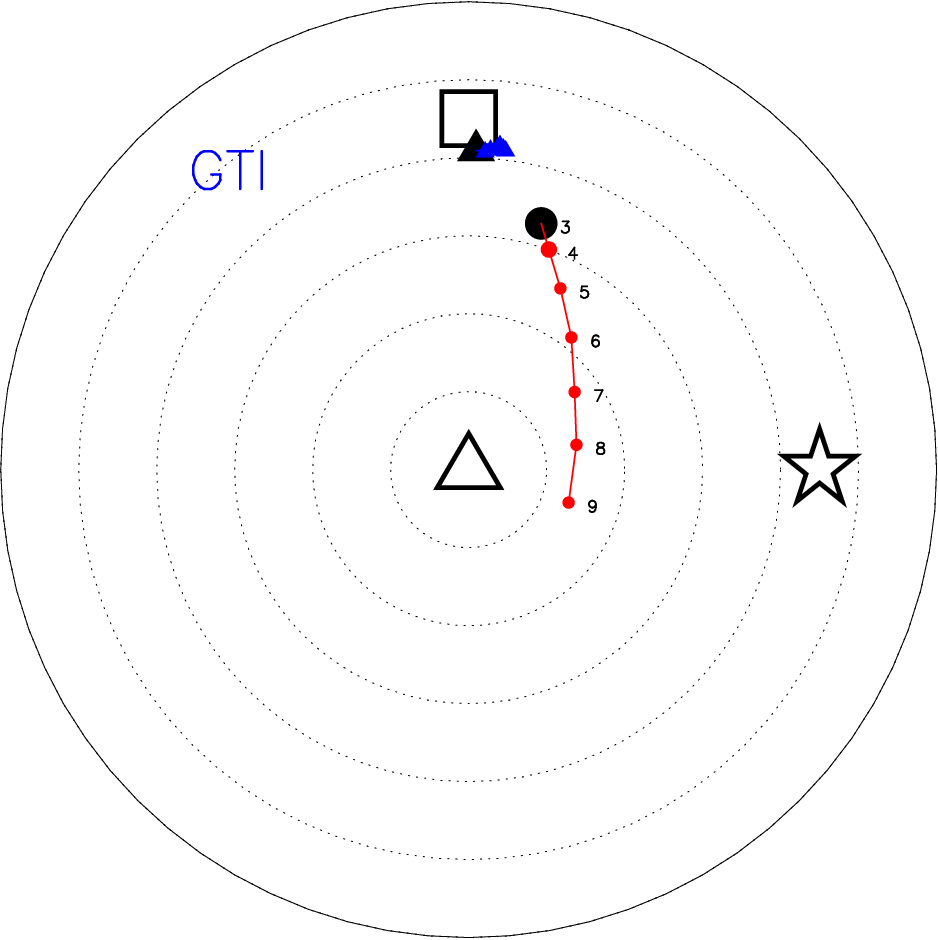}
\includegraphics[angle=0.,width=0.5\hsize]{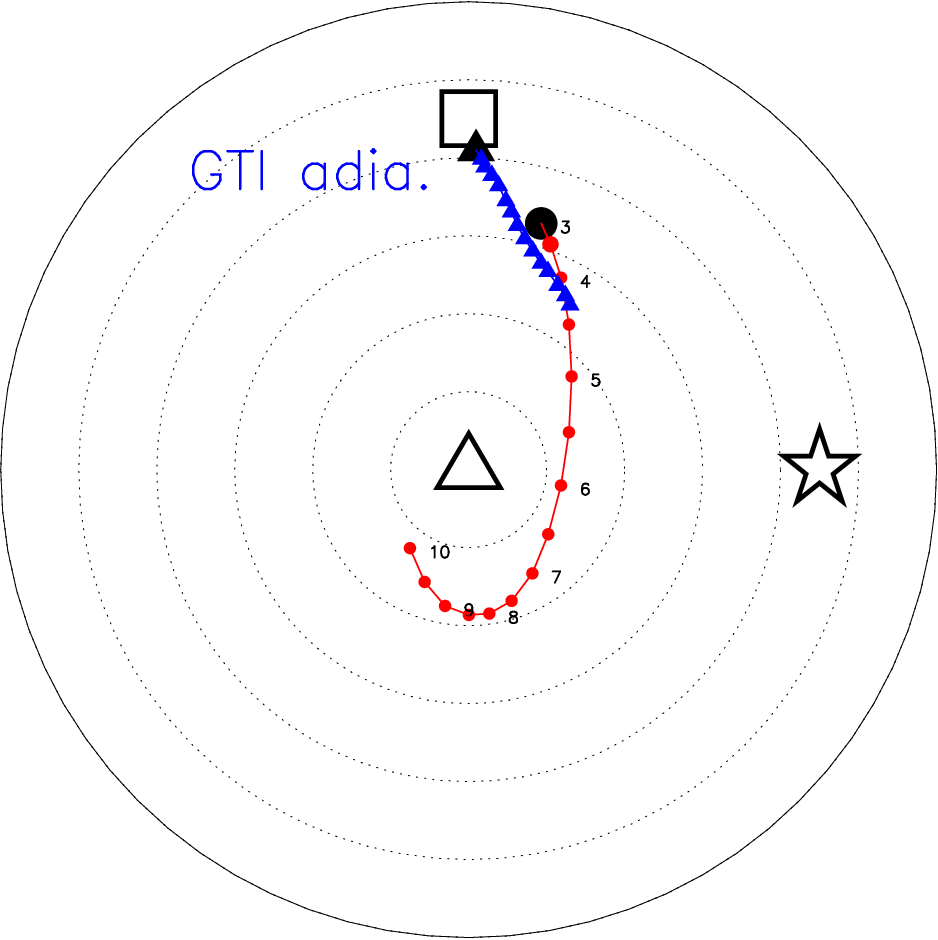}
}
\caption{ Briggs figures showing the evolution of models GTL
  (top-left), GTE (top-right), GTI (bottom-left) and GTI with star
  formation switched off after 3 Gyr (bottom-right).  Dotted circles
  are spaced by $20\degrees$.  The disc spin is indicated by the
  filled (red and black) circles.  The (blue and black) filled
  triangles mark the orientation of the hot gas angular momentum
  between 20 \kpc\ and 50 \kpc.  The spin of both the disc and gas at
  1 \Gyr\ (first output in GTL, top-left panel), 3 \Gyr\ (when the
  disc settles in model GTI, bottom-left panel and star formation
  turned off, bottom-right panel) and 2 \Gyr\ (coherent disc formed in
  GTE, top-right panel) are marked by the larger (black) corresponding
  symbols.  The long, intermediate and short axes of the haloes
  correspond to the (black) open star, square and triangle symbols.
  The outer solid circle corresponds to $\theta = 120\degrees$.}
\label{fig:gtriaxialbriggs}
\end{figure*}

\subsubsection{The Long-Axis Orientation}

The top-left panel of Fig. \ref{fig:gtriaxialbriggs} shows the
evolution of model GTL.  The disc forms nearly perpendicular to the
halo long axis and remains in this orientation for the 9 Gyr of its
evolution, in agreement with the evolution of model GPL in the prolate
halo.

\subsubsection{The Intermediate-Axis Orientation}

D13 presented the evolution of model GTI\footnote{D13 refer to this
  model as GI1.}.  Briefly, D13 showed that the disc that forms in
this model never settles into an intermediate-axis orientation, even
though the gas angular momentum is along the intermediate axis.  This
tilting can be seen in the bottom-left panel of Fig.
\ref{fig:gtriaxialbriggs}.  The failure to settle into an
intermediate-axis orientation occurs despite the fact that the inner
halo surrounding the disc is flattened vertically like the disc, so
that the disc sees only a potential with a short axis perpendicular to
it.  D13 argued that the intermediate axis orientation, which has been
suggested for the Milky Way \citep{law+09, law_majewski10,
  deg_widrow13, vera-ciro_helmi13}, is unstable {\it for the halo} and
therefore very implausible.

The bottom-right panel of Fig. \ref{fig:gtriaxialbriggs} shows the
effect of inhibiting gas cooling after 3 Gyr (by which point the disc
has already attained $\sim 50\%$ of its final mass).  Then the disc
tilts rapidly towards a short-axis orientation.  Compared with the gas
cooling case (fig. 11 of D13), the stellar disc approaches the
short-axis orientation much more rapidly when gas does not cool, with
a tilting rate almost twice that of GTI and comparable to the one in
models IA1 and IA2 without gas.

\subsubsection{Model with evolving gas angular momentum orientation}

The top-right panel of Fig. \ref{fig:gtriaxialbriggs} shows the
evolution of model GTE.  In this model the mergers are set up such
that the gas net angular momentum is initially in the plane spanned by
the long and intermediate axes.  However, torques on the corona from
the halo and angular momentum transport result in an inner ($20 \leq r
\leq 50$ kpc) gas corona with angular momentum precessing slowly about
the halo's short axis.  Between 4 and 7 Gyr this angular momentum is
within $\sim 20\degrees$ of the plane spanned by the short and long
axes of the halo, enabling the stellar disc to grow in an almost
long-axis orientation.  Indeed at 7 Gyr the disc is perpendicular to
the halo's long axis.  Yet once the gas angular momentum evolves away
from this plane, it drags the stellar disc off the long-axis
orientation by more than $20\degrees$.  That gas can drive a disc off
an equilibrium orientation demonstrates that gas cooling plays a
dominant role in determining the relative orientations of discs and
haloes.


\section{The Effect of Satellite Perturbations}

\begin{figure}
\centerline{\includegraphics[angle=0.,width=\hsize]{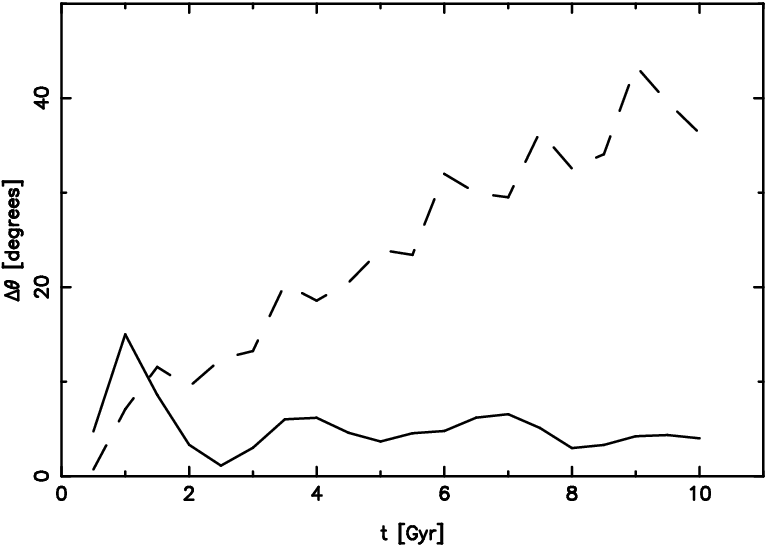}}
\centerline{\includegraphics[angle=0.,width=\hsize]{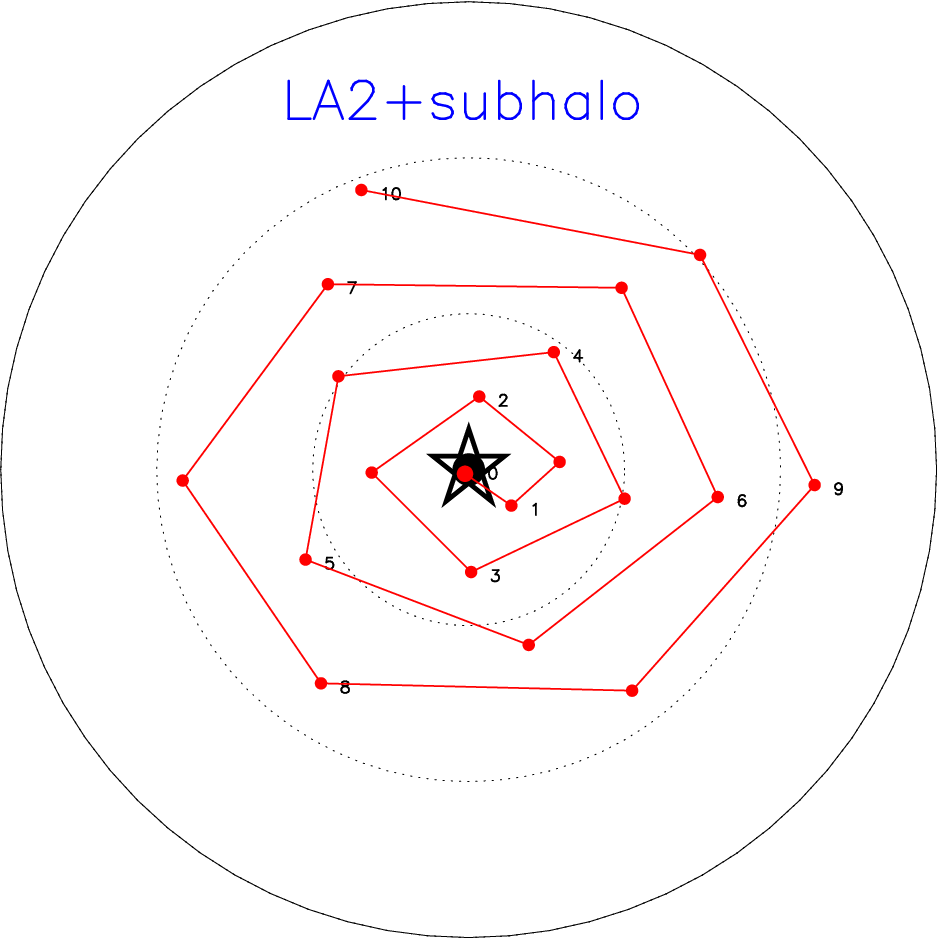}}
\caption{Top: disc tilting in models SA1 (solid line) and LA2 (dashed
  line) after the disc is perturbed by a substructure.  Plotted is the
  angle between the instantaneous angular momentum vector and the
  initial one.  Bottom: Briggs figure for model LA2 showing the
  evolution after it is perturbed by the substructure.  Dotted circles
  are spaced by $20\degrees$ with the outer solid circle corresponding
  to $\theta = 60\degrees$.  The disc spin is indicated by the filled
  (red and black) circles.  The halo long axis is indicated by the
  black open star.
  \label{fig:satperturb}}
\end{figure}

We now explore the effect of substructure on disc orientations in the
absence of gas by introducing a satellite in models SA1 and LA2.  We
use a concentration $c=12$ halo of mass $1.3 \times 10^{11} \msun$
(\ie\ $\sim 2\%$ of $M_{200}$ for halo A).  We place it at 150 kpc on
the halo's long axis and give it a tangential velocity such that its
orbital pericentre is at $\sim 10$ kpc.  In both cases, the satellite
disrupts by 2 Gyr.  The top panel of Fig. \ref{fig:satperturb} shows
the effect of this perturbation: while SA1 quickly returns to a nearly
short-axis orientation, the more massive LA2 slowly tilts away from
its original orientation for the remainder of its evolution.  These
small perturbations show that the long-axis orientation is only
quasi-stable even when the disc is massive.  The bottom panel of
Fig. \ref{fig:satperturb} shows a Briggs figure of the evolution of
the orientation of model LA2 perturbed by the satellite.  After the
initial perturbation the disc precesses about the long axis, slowly
moving away from that orientation.  This precession indicates that the
tilting is driven by the direct torquing from the halo, rather than a
lingering influence of the satellite.  The effect of the satellite
therefore is only to provide the initial perturbation that drives the
disc from the local energy minimum in the long-axis orientation.

\begin{figure*}
\centerline{
\includegraphics[angle=0.,width=0.5\hsize]{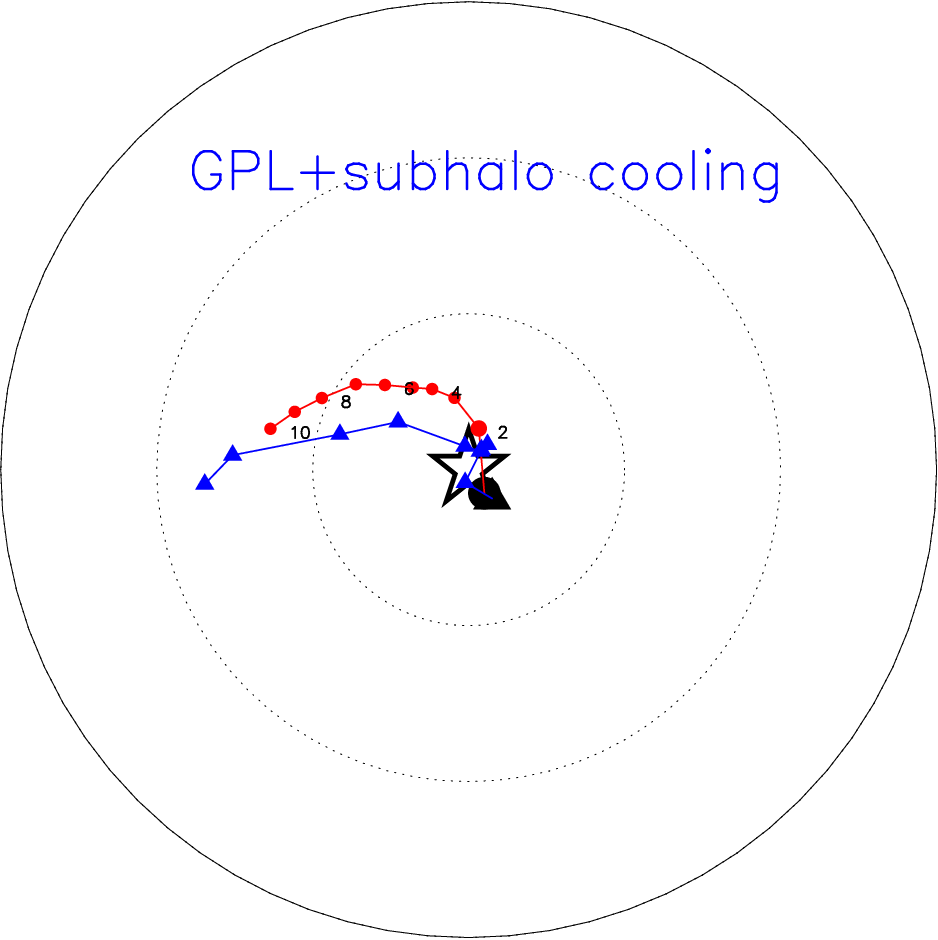}
\includegraphics[angle=0.,width=0.5\hsize]{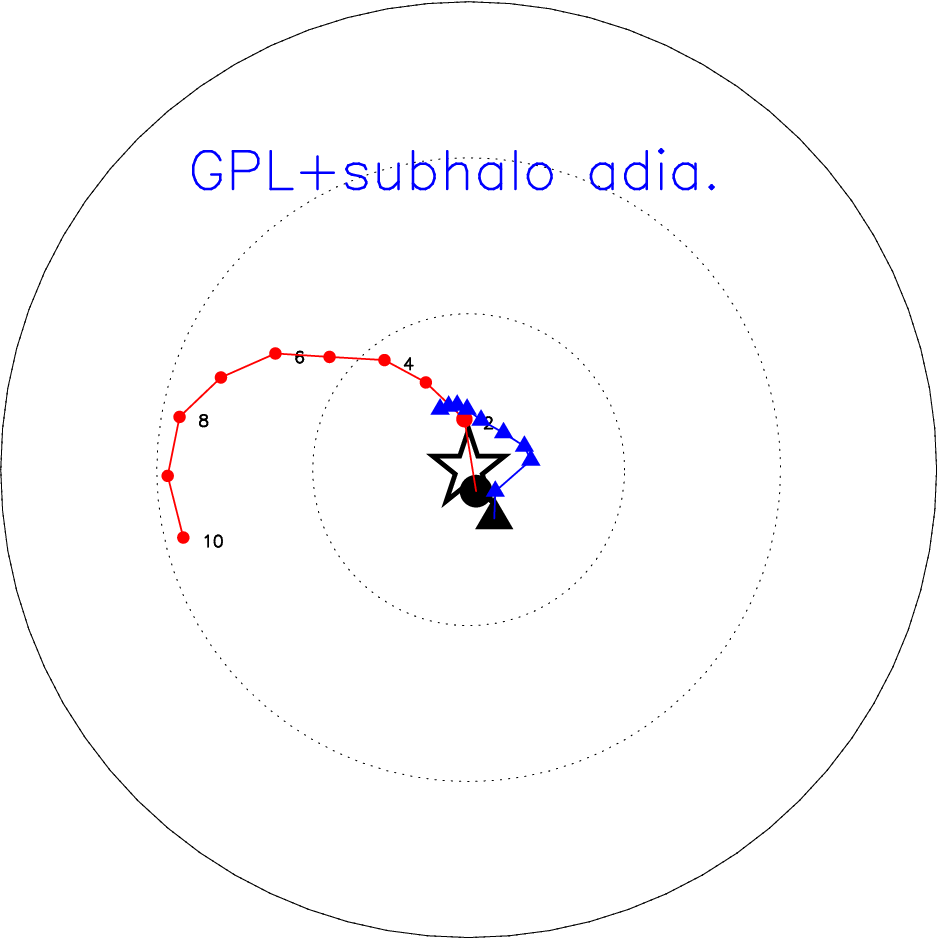}
}
\caption{ Briggs figures showing the evolution of model GPL perturbed
  by a subhalo with gas cooling and star formation (left) and evolved
  adiabatically (right).  Dotted circles are spaced by $20\degrees$.
  The disc spin is indicated by the filled (red and black) circles.
  The (blue and black) filled triangles mark the orientation of the
  hot gas angular momentum between 20 and 50 \kpc.  The spin of both
  the disc and gas at 1 \Gyr\ are marked by the larger (black)
  corresponding symbols.  The long axis of the halo is indicated by
  the (black) open star.  The outer solid circle corresponds to
  $\theta = 60\degrees$.}
\label{fig:briggs727}
\end{figure*}

To test the effect of substructure when gas is present, we insert the
same satellite into the prolate model GPL at 4~Gyr, at which point the
stellar disc has a mass of $6.6 \times 10^{10} \msun$.  The satellite
mass represents $8\%$ of $M_{200}$ for halo GP.  The satellite is
placed at 150 kpc on the halo's long axis and has a tangential
velocity such that its pericentre is at 20 kpc; the satellite disrupts
by 2 Gyr.  Fig. \ref{fig:briggs727} shows the evolution of the
system when gas is allowed to cool and when gas is adiabatic.  Both
discs tilt away from their initial orientation; the case with no gas
cooling tilts faster and further than the case with gas cooling.  In
the latter case, the disc tilting follows the angular momentum of the
gas corona, while when gas is not cooling the stellar disc tilts
independently of it.

Finite perturbations from substructures will therefore destabilize
galaxies in the long-axis orientation.  When gas cools on to the disc,
it slows down the tilting but the disc never returns to a long-axis
orientation.  Infalling satellites also impart angular momentum to the
gas corona, further driving the disc off the long-axis orientation.


\section{Discussion}
\label{sec:conc}

\subsection{Cosmological effects not considered}

\citet{valluri+13} showed that the orbital structure of a dark matter
halo surrounding a Milky Way-sized galaxy that forms in a fully
cosmological model is not much different from isolated triaxial haloes,
including some of the ones considered here.  Thus from an internal,
orbital perspective, our simulations capture the same physics as more
complicated fully cosmological simulations.

There are, however, effects which our simulations do not capture.  In
the $\Lambda$CDM universe, dark matter haloes (1) host a spectrum of
substructure accounting for roughly $3-5\%$ of the total mass of the
main halo itself at the present time \citep{diemand+07, springel+08},
(2) slowly tumble \citep{dubinski92, bailin_steinmetz04,
  bryan_cress07} (3) and gain mass.  We now consider whether each of
these differences alter our conclusions.

\subsubsection{Halo substructure}

We have shown here that satellites lead to red discs evolving to a
short-axis orientation, which is the lowest energy state of the
system.  The role of the satellites is to provide the initial finite
perturbation that allows halo torques to drive the disc to the
short-axis orientation.  For instance, a $2\%$ mass subhalo initially
drives the disc of model LA2 from the long-axis orientation by
$10\degrees$, after which the torque from the triaxial halo takes over
and drives the disc further away from the long-axis orientation.
Therefore the presence of a $\Lambda$CDM spectrum of subhaloes speeds
up the rate at which initial perturbations build up but subsequently
the evolution is dominated by the main halo itself.
\citet{kazantzidis+09} show that on average 5 subhaloes with mass
$>20\%\Mdisc$ pass within $20\kpc$ of the Galactic Centre since
redshift $z=1$.  The cumulative effect of these $\sim 10\degrees$
kicks is to ensure that the disc is never stuck in the local minimum
provided by the long-axis orientation, instead ensuring that the disc
is close to a short-axis orientation.

\subsubsection{Halo figure rotation}

Pattern rotation leads to warped planes of stable loop orbits.  The
measured pattern speeds of haloes in dark matter only simulations
follows a log-normal distribution centred on $\Omega_p = 0.148 h
\kms\kpc^{-1}$.  At these low tumbling rates, the region where stable
loop orbits are warped is at relatively large radius \citep{binney_78,
  heisler+82, magnenat_82, lake_norman_83, durisen+83,
  steiman-cameron_durisen_84, martinet_dezeeuw_88, habe_ikeuchi_85,
  habe_ikeuchi_88, deibel+11}.  \citet{deibel+11} find that as the
pattern speed gets large, both box orbits and long-axis tube orbits
are destabilized.  Prograde short-axis tubes are also destabilised,
but are replaced by retrograde short-axis tubes.  These orbital
changes can in principle alter the shapes of the haloes; however, even
the slowest of \citet{deibel+11}'s pattern speeds is about as large as
the fastest pattern speeds of haloes found in dark matter-only
simulations.  Thus, in real haloes loop orbits at small radii should
still be stable circulating about the short and long axes.  The
short-axis orientation should therefore still be a stable orientation
for stellar discs.  We therefore conclude that the tumbling of dark
matter haloes is not large enough to alter our conclusions.

\subsubsection{Halo growth}

Haloes are generally believed to be continuously growing.  For
instance the virial radius and concentration of haloes increase even
at late times.  However much of this is pseudo-evolution due to the
changing critical density of the universe, relative to which the
virial radius and concentration parameters are defined.
\citet{diemer+13} show that for galaxy-sized haloes, $M_{200} \sim
10^{12} \msun$ at $z=0$, most of the evolution since $z = 1$ can be
attributed to pseudo-evolution of this kind.  Likewise,
\citet{prada+06} find little evolution in the outer regions of haloes
in the mass range $10^{11}$-$5\times10^{12}\msun$.  Thus, most disc
galaxies will not experience significant halo growth.  More massive
haloes, which typically host elliptical galaxies, will still be
growing rapidly and these will drive evolution not considered here.

\subsubsection{Halo axes-ratios}

One important difference between our simulations and cosmological ones
is in the shapes of the haloes.  The haloes in our simulations are
generally quite elongated, with $0.32 \la c/a \la 0.55$; less than
$3\%$ of haloes of mass $\log M_{200} \simeq 12.5$ in cosmological
simulations have $c/a < 0.32$, while $>16\%$ of haloes have $c/a <
0.55$ \citep{maccio+07b, maccio+08, schneider+12}.  Therefore, the
torques from the dark matter haloes that the discs in the simulations
experience are larger than what might be expected in general.  This
has the effect that it allows gasless discs to settle into a
minor-axis orientation faster here than in nature.  On the other hand,
gas cooling is able to tilt disc galaxies even more readily than in
our simulations.

We conclude that while our simulations are idealized, they capture
most of the physics relevant for the relative orientation of discs and
haloes.  The elongated shapes of our haloes may bias the galaxies into
stronger alignment with their haloes, but we already find that gas
dominates the disc alignment in the case that gas cools on to the
disc, which makes the gas even more dominant when the haloes are
rounder than here.  \citet{debuhr+12} presented cosmological
simulations in which stellar discs were placed with their spin along
either the short or the long axes of their haloes.  They found that
discs tumble significantly, particularly when they start out in the
long-axis orientation, but generally remain aligned with one of the
main axes inside 50 kpc.  These results with a cosmological population
of subhaloes and realistic halo tumbling rates are consistent with our
findings, confirming that satellites destabilize discs in a long-axis
orientation but do not alter the orientation of discs in the
short-axis orientation quite so much.

\subsection{Red galaxies}

In the absence of gas cooling, the main driver of a disc's orientation
relative to its triaxial halo is the stability of different
orientations.  Loop orbits are stable around the short and long axes,
but not around the intermediate axis \citep{hei_sch_79, schwar_79, 
  goo_sch_81, wil_jam_82, ada_etal_07, carpintero_muzzio11}.  This has
led to the expectation that discs spin around either the long or the
short axis of a triaxial halo.

In the absence of gas, the orientation in which the disc's spin and
halo's short axis align is always stable.  Indeed this orientation has
the lowest potential energy.  We show that a disc can never remain
with its spin aligned with the intermediate axis of the halo, even if
the disc flattens the inner halo so much that the vertical direction
is the halo's shortest axis throughout the disc.  The equilibrium of
discs orthogonal to the long axis of the halo is unstable, but the
tilting rate is so low for massive discs that the orientation persists
for a Hubble time and can be considered quasi-stable.  However,
perturbations from quite low-mass and low-concentration satellites
will drive even massive discs away from this orientation, whereas a
disc with its spin along the halo's short axis returns to that
orientation soon after a satellite perturbs it.  

Red galaxies will therefore tend to line up with their spin parallel
to the short axis of the halo.  Indeed studies using the Sloan Digital
Sky Survey have found that red galaxies have a satellite distribution
which is aligned with their projected major axis
\citep{sales_lambas04, brainerd05, agustsson_brainerd06, yang+06,
  bailin+08}.  Alignment is not however perfect, because discs do not
settle into a short-axis orientation immediately.

We find that, at fixed mass, a disc in the long-axis orientation tilts
much more rapidly when it has a larger scalelength.  This occurs
because more of the disc is exposed to the part of the potential that
is perpendicular to the disc.  A testable prediction of our
simulations therefore is that, at fixed mass, red galaxies with larger
half-mass radius should be more tightly coupled to their haloes and
therefore show a stronger alignment.

\subsection{Blue galaxies}

The continued star formation activity of blue galaxies requires the
accretion of gas \citep[e.g.][]{sancisi+08}.  Cosmological simulations
find that gas and dark matter angular momentum decoupling is common
\citep[e.g.][]{bailin+05, croft+09, bett+10, hahn+10, roskar+10}.  Our
simulations show that, in order for the stellar disc to reach a
balance between the gravitational torques and the angular momentum it
gains from the halo the gas spin must in general be misaligned with
the disc's, regardless of whether the gas is shock heated, or arrives
via cold flows \citep{bir_dek_03, ker_etal_05, dek_etal_09}.  This
leads to star-forming blue discs remaining in almost arbitrary
orientation.  Rather than being perpendicular to the gas angular
momentum, the disc tends to align itself at an orientation where the
torque from the halo and the gas inflow balance.  The supply of gas is
therefore a much stronger driver of disc orientation than are
gravitational torques from the halo, even when the halo globally has a
quite prolate-triaxial shape.  When the gas corona angular momentum
orientation is evolving, the disc orientation also changes to remain
in equilibrium between the halo torque and the inflowing gas angular
momentum.  

\citet{fraternali_tomassetti12} estimate a star formation rate of
$\sim 3 \msun \mathrm{yr}^{-1}$ in the Milky Way.  Assuming a Milky
Way disc mass of $4$-$5 \times 10^{10} \msun$, this corresponds to
specific star formation rate (SSFR) $\dot M_*/M_*\sim 0.06-0.08
\Gyr^{-1}$.  While GP45 has a high star formation rate, $\sim 7 \msun
\mathrm{yr}^{-1}$, the corresponding SSFR is $\dot M_*/M_*\sim 0.08
\Gyr^{-1}$, in good agreement with the Milky Way.  Thus, the SSFR, and
presumably the cooling rates, in the simulations are characteristic of
real galaxies and we can expect that galaxies like the Milky Way will
be off the main planes of their dark matter haloes.  At lower gas
cooling rates, and lower SSFRs, gas cooling is no longer able to
deliver sufficient angular momentum to the disc and the disc drops
into a short-axis alignment with the halo.

Even in the presence of cooling gas the intermediate-axis orientation
continues to be avoided.  Since the distribution of orientations
avoids one axis, this introduces a weak correlation between the halo
long axis and disc spin.  D13 found that the stellar disc that forms
when gas angular momentum is along the intermediate axis is never
closer to it than $20\degrees$.  
Detailed modelling, which will be presented elsewhere, suggests that
this tendency to avoid alignment along the intermediate axis results
in a weak {\it anti-correlation} between the major axis of the disc
and the projected major axis of its host halo. The observed, weak
alignments for blue galaxies therefore either require a correlation
between the gas angular momentum and the large scale structure, or a
contribution from galaxies in which the gas inflow rate is
insufficient to (significantly) impact the orientation of the disc.

The observed difference in relative orientation of disc and halo
between red and blue galaxies is additional evidence that star-forming
galaxies need to continuously accrete gas to maintain their star
formation.  Without the inflow of misaligned angular momentum, torques
from triaxial haloes and perturbations from subhaloes would drive the
discs of blue galaxies to an alignment similar to that of red
galaxies.  The important difference between blue and red galaxies is
not the presence of gas and star formation but the cosmic infall of
gas with misaligned angular momentum with some of this gas cooling
on to the stellar disc.  A galaxy slowly forming stars from only a
pre-existing gas disc/inner corona (including the recycling of gas via
feedback and a fountain) will still tend to evolve towards a
short-axis orientation.

\subsection{The Milky Way and Andromeda}

The shape of the Milky Way's dark matter halo has been modelled
numerous times via the Sagittarius stream.  To date the best-fitting
models have needed a triaxial halo, but have required that the disc is
perpendicular to the intermediate axis of the halo \citep{law+09,
  law_majewski10, deg_widrow13, vera-ciro_helmi13}.  D13 showed that
this orientation is particularly unstable and very unlikely even if
the inner halo is flattened.  By a process of elimination, they
concluded that the Milky Way must be tilted with respect to the
principal planes of the halo.  \citet{deg_widrow14}, using mock data
of a disc tilted relative to the halo, showed that modelling under the
incorrect assumption that the disc lies in one of the main planes of
the halo leads to a best-fitting model with the disc perpendicular to
the intermediate axis of the halo.  This may explain the persistent
finding that the Milky Way is perpendicular to its halo's intermediate
axis.

The case for a tilted disc in Andromeda is suggested by the
distribution of a fraction of the satellites around it, which
\citet{ibata+13b} showed consists of a thin disc inclined relative to
the main plane of the stellar disc.  \citet{bowden+13} show that the
satellite disc must lie perpendicular to the long or the short axis of
the halo if the satellite disc is long lived and the halo is not
spherical, suggesting that the main stellar disc is inclined relative
to the halo.

\subsection{Summary}

Our main results can be summarized as follows.

\begin{itemize}

\item A stellar disc without gas can persist indefinitely with its
  spin along the short axis of a triaxial halo.  Even when perturbed
  by a satellite, the disc quickly settles back to this orientation.
  Instead if a stellar disc starts with its spin aligned with the long
  axis of the halo, whether the disc remains in this orientation or
  not depends on the shape of the halo potential.  At low stellar mass,
  the potential surrounding the disc is orthogonal to it and the disc
  tilts towards a short-axis orientation.  On the other hand, when the
  disc is massive it flattens the total potential and it usually can
  persist in this orientation for a long time.  Perturbations by
  satellites however permanently drive a disc off the long-axis
  orientation.  Finally, not even a very massive stellar disc is
  stable perpendicular to the intermediate axis of the dark matter
  halo, even when the inner halo becomes flattened.  Thus, the most
  natural orientation for gasless red discs is with their spins along
  the short axis of the halo.

\item A disc with gas cooling on to it settles into equilibrium between
  the angular momentum it gains from the gas and the torques it feels
  from the halo and gas corona.  The disc is therefore not
  perpendicular to the axes of the halo (unless the gas spin is about
  the short or the long axis) nor is it perpendicular to the spin of
  the gas.  If the orientation of the gas angular momentum evolves,
  then the orientation of the disc is also forced to change.

\item These simulations therefore imply that while gas poor, red
  galaxies tend to align with their spins along the short axis of
  their haloes, the infall of gas on to blue galaxies allows them to
  linger at a wide range of orientations relative to the halo.  Thus,
  stacking blue haloes lead to a nearly isotropic distribution.  The
  need to avoid the intermediate-axis orientation introduces a weak
  anticorrelation between the disc and the halo major axis.

\item Our simulations show that the relative orientations of discs and
  haloes are set by local conditions, \ie\ those within the virial
  radius of the halo.

\end{itemize}

\bigskip
\noindent
{\bf Acknowledgments.} 

\noindent We thank the anonymous referee for a report that helped to
improve this paper.  Collisionless simulations in this paper were
carried out on the Arctic Region Supercomputing Center as well as at
the DIRAC Shared Memory Processing system at the University of
Cambridge, operated by the COSMOS Project at the Department of Applied
Mathematics and Theoretical Physics on behalf of the STFC DiRAC HPC
Facility (www.dirac.ac.uk). This equipment was funded by BIS National
E-infrastructure capital grant ST/J005673/1, STFC capital grant
ST/H008586/1, and STFC DiRAC Operations grant ST/K00333X/1. DiRAC is
part of the National E-Infrastructure.  Simulations with gas were
carried out using computational facilities at the University of Malta
procured through the European Regional Development Fund, Project
ERDF-080\footnote{http://www.um.edu.mt/research/scienceeng/erdf\_080)},
at the Texas Advanced Computing Center
(TACC)\footnote{http://www.tacc.utexas.edu} at The University of Texas
at Austin, on Kraken at the National Institute for Computational
Sciences\footnote{http://www.nics.tennessee.edu/} and at the High
Performance Computer Facility of the University of Central Lancashire.
VPD and DRC are supported by STFC Consolidated grant \#~ST/J001341/1.
VPD thanks Jairo M\'endez-Abreu for useful discussions.

\bigskip 
\noindent

\bibliographystyle{aj}
\bibliography{ms.bbl}

\label{lastpage}

\end{document}